\documentclass[%
 reprint,
 amsmath,amssymb,
 aps,
onecolumn
]{revtex4-2}

\usepackage{graphicx}
\usepackage{dcolumn}
\usepackage{bm}
\usepackage{mathtools}
\usepackage{bbold}
\usepackage{subfig}
\usepackage{comment}

\begin{document}
\global\long\def\Nact{M}
\global\long\def\U{\mathrm{U}}
\global\long\def\r{\boldsymbol{r}}
\global\long\def\Jspace{K}
\global\long\def\Jdisord{J}
\global\long\def\Atwopoint{K_{\mathrm{rec}}}
\global\long\def\wtwopoint{w_{\mathrm{rec}}}
\global\long\def\Aonepoint{\U_{\mathrm{inp}}}
\global\long\def\wonepoint{w_{\mathrm{inp}}}

\global\long\def\Tc{\mathrm{T}}
\global\long\def\ro{\boldsymbol{w}}

\global\long\def\kv{v}
\global\long\def\ks{\kappa^{3}}
\global\long\def\kk{\kappa^{4}}
\global\long\def\kvvec{\boldsymbol{v}}
\global\long\def\ksvec{\boldsymbol{\kappa}^{3}}
\global\long\def\Kv{V}

\global\long\def\bphi{\boldsymbol{\phi}}
\global\long\def\bpsi{\boldsymbol{\psi}}

\preprint{APS/123-QED}

\title{Information content in continuous attractor neural networks is preserved in the presence of moderate disordered
background connectivity}


\author{Tobias K\"uhn$^{1,2}$}
\email{tobias.kuhn@inserm.fr}
\author{R\'emi Monasson$^1$}
\affiliation{%
$^1$ Laboratoire de Physique de l'Ecole Normale Sup\'erieure, CNRS UMR8023 \&\\
PSL Research, Sorbonne Universit\'e, Universit\'e Paris Cit\'e, F-75005 Paris, France \\
$^2$ Institut de la Vision, Sorbonne Universit\'e, CNRS, INSERM, F-75012 Paris, France 
}%

\date{\today}

\begin{abstract}
Continuous attractor neural networks (CANN) form an appealing conceptual model for the storage of information in the brain. However a drawback of CANN is that they require finely tuned interactions. We here study the effect of quenched noise in the interactions on the  coding of positional information within CANN. Using the replica method we compute the Fisher information for a network with position-dependent input and recurrent connections composed of a short-range (in space) and a disordered component. We find that the loss in positional information  is small for not too large disorder strength, indicating that CANN have a regime in which the advantageous effects of local connectivity on information storage outweigh the detrimental ones. Furthermore, a substantial part of this information can be extracted with a simple linear readout. 
\end{abstract}

\maketitle


\section{\label{sec:intro}Introduction}
The ring attractor neural network was proposed by Amari in the 70's as a practical  way to memorize a collective variable within a noisy neural population \cite{Amari77}. This work opened the way to various theoretical applications of the concept of continuous attractor neural networks (CANN), {\em e.g.} in the contexts of the orientational tuning \cite{Benyishai95} or  hippocampal place cells \cite{Tsodyks95}, as well as to extensions, in particular to the case of multiple attractor embeddings \cite{Battaglia98,romani10,Battista20}. While indirect evidence for the existence of CANN could be found in recordings of activity in the hippocampus \cite{posani18}, in the enthorinal \cite{Yoon2013} and the prefrontal cortex \cite{Wimmer2014}, a direct and beautiful observation of ring attractor coding for head direction was obtained only recently in the ellipsoid body of the fly \cite{Kim17_849}. 

From a theoretical point of view, CANN models rely on recurrent excitatory interactions between neurons active for similar values of the encoded variable, {\em e.g.} the position of the animal in physical space, together with a long-range inhibition preventing all cells to be active together. This combination of local positive interactions and global inhibition creates a localized bump of activity, whose center of mass reliably represents the collective variable. In this regard, a crucial condition is that the bump can be easily moved (under weak external, "sensory", inputs) to span the continuous set of values of the variable. This condition imposes that the short-range connections are finely tuned, so that the model be effectively translation invariant.

When the finite-tuning condition breaks down, {\em e.g.} due to random modulations of the interactions, the bump can get stuck in the absence of neural noise \cite{Tsodyks95}.  In practice, quenched noise in the interactions can come from imperfect learning of one environment, or from interferences resulting from other information encoded (maps, objects distorting the map locally, ...). Quantifying the loss in the accuracy of information storage resulting from heterogeneities in the interactions is an important issue.

We address this question here in the framework of decoding of information, based on analytical and numerical calculations. We propose an analytically tractable model of binary (active/silent) neurons receiving position-dependent inputs, and connected to each other through spatially coherent and short-range interactions, on top of a disordered and incoherent background. Using the replica method we compute the Fisher information in the high-dimensional neural activity about the encoded position as a function of the intensity of disordered interactions. 
This quantity was identified as a measure that is both relatively easy to compute for many systems and objectively quantifies the information contained in the neural activity about the stimulus (an orientation or a point in space) \cite{Seung93_10749}.
It is more appropriate for this quantification than, for example, the readout of the center of mass of a bump of activity \cite{Pouget99_85}. Yet, the Fisher information is not an information measure in the sense of Shannon. From this point of view, the mutual information between the stimulus and the neural activity is the quantity we are eventually interested in \cite{Cover91}. However, the latter is a global quantity, integrated over all possible stimuli and its computation is generally more difficult than that of the Fisher information, which puts restrictions on the system it can be calculated for \cite{Sompolinsky01_051904}. 
In the thermodynamic limit, the mutual information can be obtained from the Fisher information under the condition that the correlations are not too strong \cite{Brunel98_1731}. We have explicitly checked that this prerequisite is fulfilled by our model, so that the computation of the Fisher information is actually sufficient. For a current review discussing both measures of information and their use in neuroscience, see e.g. \cite{Kriegeskorte21_703}.
\\
The paper is organized as follows. In sec. \ref{sec:methods},  our model is introduced. We establish the Fisher information as a means to quantify the information contained in the neural activity about the stimulus, together with its relation to other information-theoretic measures, compute it in the thermodynamic (mean-field) limit and derive its analytical properties in the limiting case of weak connection strengths. In sec. \ref{sec:Validation_Application}, we validate our mean-field results by means of Monte-Carlo calulations, study the dependence of the Fisher information on changes in the recurrent and feed-forward connectivity and how precise a linear readout can be compared to the bound predicted by the Fisher information. In sec. \ref{sec:Discussion}, we put our results into context and give an outlook to possible future directions.

\section{Model and methods\label{sec:methods}}
\begin{figure}
	\includegraphics[width=.5\textwidth]{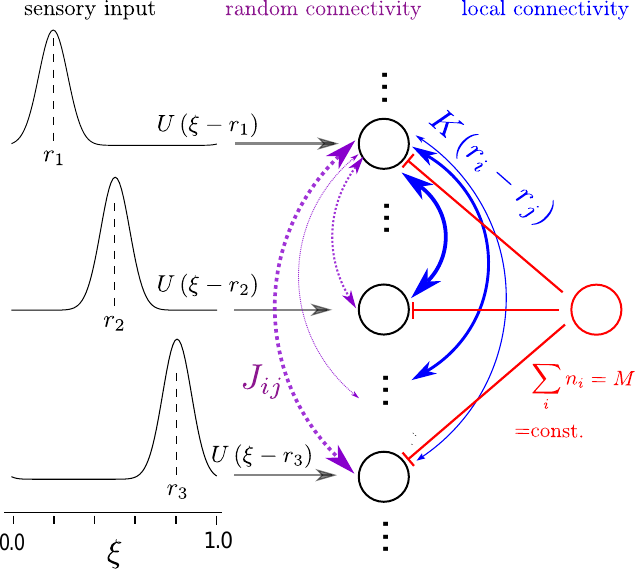}
	\caption{Scheme of the network model employed in this work. The bell-shaped curves represent the space sensitive single-neuron (feedforward) input and the red circle a symbolic inhibitory neuron that ensures that the summed activity of all neurons is constant.}
	\label{fig:Network_scheme}
\end{figure}

\subsection{Distribution of neural activities}

We model neurons as binary units, taking the value $1$ when active and $0$ when inactive. Each neuron is receiving a `sensory' input, whose value depends on the mismatch between the position $r_i$ in physical space it is maximally responding to and the position of the 'animal', see fig. (\ref{fig:Network_scheme}) for a scheme of our setup. The probability distribution of activities is governed by a Boltzmann law $P\left(\mathbf{n}|\xi\right) \sim e^{-E\left[\boldsymbol{n}\right]}$ and the energy
\begin{align}
	E\left[\boldsymbol{n}\right] = -\frac{1}{2}\sum_{i\neq j}\left(\Jdisord_{ij}+\Jspace_{ij}\right)n_{i}n_{j}-\sum_{i}n_{i}\, \U\left(\xi-r_{i}\right),
\end{align} 
where $\Jspace_{ij} = \Jspace \left(r_{i} - r_{j}\right)$ is the local part of the interaction that we assume to be decaying in space with a typical length scale $\wtwopoint$ and strength $\Atwopoint$. $\Jdisord$ is the disordered part of the connectivity with the statistics
\begin{align}
    \left\langle J_{ij}\right\rangle = 0,\ \left\langle J_{ij}J_{i^{\prime}j^{\prime}}\right\rangle =\frac{g^{2}}{2N}\delta_{ii^{\prime}}\delta_{jj^{\prime}}\label{eq:statistics_disorder}
\end{align}
and $\U$ mimics the space-dependent input, which we model by
\begin{align}
	\U\left(\Delta x\right) = \Aonepoint  \exp\left( - \frac{\Delta x^{2}}{\wonepoint^{2}}\right).
\end{align}
We are assuming periodic boundary conditions. As a simple way to model global inhibition, we impose the constraint that the summed activity is fixed to $\sum_{i} n_{i} = f\cdot N = M$, where $f\in\left[0,1\right]$.
Our model is closely related to that of \cite{Rosay13_062813}, where the case of a proper continuous attractor neural network (CANN) is studied, so the connectivity matrix is composed of a sum of local connectivities in different environments. 
However, in this study we are interested in the effect of the disorder on the information content in the neural activity for a single environment. Also, the presence of other maps is not the only source of disorder as there is always some variability in the connectivity. Therefore, to simplify the setup, we content ourselves with approximating the disordered contribution to the connections as Gaussian, which also corresponds to the high-temperature behaviour of the disorder in \cite{Rosay13_062813}. 

\subsection{\label{sec:FisherInfoIntro}Fisher information and mutual information}
We now want to quantify the amount of information contained in the neural activity. One possibility to do so is to compute the Fisher information for a given stimulus $\xi$,
\begin{equation}
    {\cal I}_{\boldsymbol{n}}\left(\xi\right) \coloneqq \left\langle -\frac{\partial^{2}}{\partial \xi^{2}} \ln P\left(\boldsymbol{n}|\xi\right)\right\rangle_{\boldsymbol{n}},
    \label{eq:Def_FisherInfo}
\end{equation}
a standard measure for the quantification of information in neural populations \cite{Pouget99_85, Kriegeskorte21_703}. According to the Cram\'er-Rao bound, its inverse gives a lower bound on the variance of any unbiased estimator of $\xi$ \cite{Cover91}. We will discuss this relation in greater depth in section \ref{sec:Linread}.
Furthermore, in the thermodynamic limit, that we are interested in, it also determines the mutual information, a connection first established in \citep{Brunel98_1731} and later refined in \cite{Wei16_305}. The mutual information is given by the decrease of the entropy of the neural activity due to the knowledge of the stimulus, concretely
\begin{align*}
	I_{\mathrm{MI}} &\coloneqq - \sum_{\boldsymbol{n}} P\left(\boldsymbol{n}\right) \ln P\left(\boldsymbol{n}\right) + \int d\xi\, p\left(\xi\right) \sum_{\boldsymbol{n}} P\left(\boldsymbol{n}|\xi\right) \, \ln P\left(\boldsymbol{n}|\xi\right)\\
	\mathrm{where} \ P\left(\boldsymbol{n}\right) &\coloneqq \int d\xi\, p\left(\xi\right) P\left(\boldsymbol{n}|\xi\right).  
\end{align*}
In \citep{Brunel98_1731} (their eq. (13)), the relation
\begin{align}
	I_{\mathrm{MI}} = -\int d\xi \,p\left(\xi\right) \ln p\left(\xi\right)
			+ \int d\xi\,  p\left(\xi\right) \ln\left(\frac{{\cal I}_{\boldsymbol{n}}\left(\xi\right)}{2\pi e}\right) + \mathcal{O}\left(\frac{1}{N}\right) 
\end{align} 
was derived for an ensemble of neurons with fixed covariances, without disorder. However, in this study we are limiting ourselves to the saddle-point approximation of the Fisher information, which is valid up to corrections of order $1/N$ as well. So the presence of disorder does not change much and we obtain that
 \begin{align}
	\left\langle I_{\mathrm{MI}}\right\rangle_{J} = -\int d\xi \, p\left(\xi\right) \ln p\left(\xi\right)
			+ \int d\xi\,  p\left(\xi\right) \ln\left(\frac{\left\langle {\cal I}_{\boldsymbol{n}}\left(\xi\right)\right\rangle_{J}}{2\pi e}\right) + \mathcal{O}\left(\frac{1}{N}\right),\label{eq:Relation_Fisher_mutual_disordered} 
\end{align} 
where $\left\langle \dots \right\rangle_{J}$ is indicating the average over the disordered connectivity $J$. In appendix \ref{sec:MutualInfo_FisherInfo_independent}, we rederive this relation, for unconnected neurons, but more directly than in \cite{Brunel98_1731}.\\
Hereafter, we will focus on the Fisher information, which is easier to obtain than the mutual information that we get for free due to eq. (\ref{eq:Relation_Fisher_mutual_disordered}). 
Determining the Fisher information for our model, we obtain from eq. (\ref{eq:Def_FisherInfo}) after some lines of computation, detailed in appendix \ref{sec:CompFisherInfo},
\begin{align}
	\label{eq:Def_disordered_FisherInfo}
	{\cal I}_{\boldsymbol{n}}\left(\xi\right) = &\sum_{i,j}\U^{\prime}\left(\xi-r_{i}\right)\left\langle \left[\left\langle n_{i}n_{j}\right\rangle _{\boldsymbol{n}}-\left\langle n_{i}\right\rangle _{\boldsymbol{n}}\left\langle n_{j}\right\rangle _{\boldsymbol{n}}\right]\right\rangle _{\Jdisord}\U^{\prime}\left(\xi-r_{j}\right)\\
	= & \left[\boldsymbol{\U}^{\prime}\left(\xi-\r\right)\right]^{\mathrm{T}}C\boldsymbol{\U}^{\prime}\left(\xi-\r\right),
\end{align}
where $C$ denotes the disorder-averaged covariance matrix of $\boldsymbol{n}$. Conditioned on one realization of the disorder furthermore, we have introduced the thermal average
\begin{align}
	\left\langle f\left(\boldsymbol{n}\right)\right\rangle _{\boldsymbol{n}}:=\frac{1}{{\cal Z}_{\Jdisord}\left(\xi\right)}\sum_{\boldsymbol{n}}f\left(\boldsymbol{n}\right)e^{-E\left[\boldsymbol{n}\right]},
\end{align}
for some function $f$, together with  the partition function
\begin{align}
	{\cal Z}_{J}\left(\xi\right) \coloneqq \sum_{\boldsymbol{n}}e^{-E\left[\boldsymbol{n}\right]}.
\end{align}

Eq. (\ref{eq:Def_disordered_FisherInfo}) can be brought into a more familiar form by noting that
\begin{align}
	\frac{\partial}{\partial \xi} \Tc \left(\xi - r_{i}\right)&\coloneqq \frac{\partial}{\partial \xi} \left\langle n_{i} \right\rangle _{\boldsymbol{n}} = \left\langle n_{i} \left[\sum_{j} \U^{\prime}\left(\xi - r_{j}\right) \left(n_{j} - \left\langle n_{j}\right)\right\rangle \right]\right\rangle _{\boldsymbol{n}}
	 = \left(C \boldsymbol{\U}^{\prime}\right)_{i} \\
	 \Leftrightarrow \U^{\prime}\left(\xi - r_{i}\right) &= \left(C^{-1} \boldsymbol{\Tc^{\prime}}\right)_{i},
\end{align} 
where we have introduced the tuning curve $\Tc$ of neuron $i$ indicating its average activity given the input $\xi$. With this, the Fisher information can be written as \cite{Sompolinsky01_051904}
\begin{equation}
    {\cal I}_{\boldsymbol{n}} \left(\xi\right) = \left[\boldsymbol{\Tc}^{\prime}\left(\xi\right)\right]^\mathrm{T} C^{-1} \boldsymbol{\Tc}^{\prime}\left(\xi\right).
    \label{eq:FisherInfo_with_tuning_curves}
\end{equation}
This form is more handy when dealing with experimental data because the tuning curve is (in principle) a directly measurable quantity, whereas $\U^{\prime}$ is not. For our purposes, however, the form of eq.  (\ref{eq:Def_disordered_FisherInfo}) is more practical because there, the only quantity depending on the disorder is the covariance matrix. We therefore only have to compute the disorder average of the covariance matrix, which we will tackle in the following.
\subsection{Disorder-averaged statistics}
As usual for disordered systems \cite{Mezard87}, we determine the statistics from the logarithm of the partition function, the cumulant-generating functional (or Gibbs free energy)
\begin{align}
	\label{eq:Def_disorder_averaged_cumulant_gen_fct}
	\left\langle W\left(\boldsymbol{h}\right)\right\rangle_{J}= & \int dJ\,P\left(J\right)\ln\left[\sum_{\boldsymbol{n},\, \sum_{i} n_{i} = M}e^{\frac{1}{2}\sum_{i\neq j}\left(\Jdisord_{ij}+\Jspace_{ij}\right)n_{i}n_{j}+\sum_{i}n_{i}\left[\U\left(\xi-r_{i}\right)+h_{i}\right]}\right].
\end{align}
The computation of $W$ proceeds along the classical lines \cite{Sherrington75_1792}, with the difference that we have a local connectivity. We therefore do not only introduce the Gaussian helping field $q$ to decouple the four-point terms emerging from the disorder average, but also a space-dependent (also Gaussian) order parameter $\phi_{x}$ to decouple the local term $\boldsymbol{n}^{\mathrm{T}} K \boldsymbol{n}$. As apparent from the saddle-point equations (\ref{eq:Saddle_point_q_main}) and (\ref{eq:Saddle_point_phi_final_main}), $q$ quantifies the population-averaged variance of the activity and $\phi_{x}$ the population-averaged input to the neuron with place field at position $x$. Furthermore, due to the restriction on the summed activity, we introduce the Lagrange multiplier $\lambda$. As derived in appendix \ref{sec:CompFisherInfo}, we obtain the disorder-averaged cumulant-generating function in the thermodynamic limit $N\rightarrow \infty$
\begin{align}
    \left\langle W\left(\boldsymbol{h}\right)\right\rangle _{J}
= & \underset{q,\bar{q}, \bpsi, \bphi, \lambda}{\mathrm{extr}} \left\{\frac{1}{2}Ng^{2}q^{2}-\frac{1}{2}Ng^{2}\bar{q}^{2}-\frac{1}{2}\bphi^{T}K^{-1}\bphi - N \left(\lambda - g^{2}\left(\bar{q} - q\right)\right) f.\right.\\
 & \left.+\prod_{y}\frac{1}{\sqrt{2\pi}}\int dt_{y}\,e^{-\frac{t_{y}^{2}}{2}} \sum_{x}\ln\left[1+e^{\phi_{x}+t_{x}g\sqrt{2q}+\U\left(\xi-r_{x}\right)+ \lambda + h_{x}}\right]\right\}\\
 \eqqcolon & G\left(\boldsymbol{h}, \boldsymbol{\phi}, q\right),
 \label{eq:Cumulant_generating_functional_thermodynamic_limit}
\end{align}
where the "extr." implies a supremum over $q$, $\bar{q}$, $\bpsi$ and $\bphi$ and an infimum over $\lambda$. We comment on the latter point in appendix \ref{sec:App_fixedAct_spacedep_coupl}. As detailed in appendix \ref{app:disorder_saddle_point}, we obtain the saddle-point equations
\begin{align}
     q & =\int dx\,\int Dt_{x}\,\frac{1}{\left[1+e^{-\left(\phi_{x}+t_{x}g\sqrt{2q}+\U\left(\xi-x\right) + \lambda\right)}\right]^{2}},\label{eq:Saddle_point_q_main}\\
   \phi_{x} & =\int dy\, K\left(x-y\right)\int Dt_{y}\,\frac{1}{1+e^{-\left(\phi_{y}+t_{y}g\sqrt{2q}+\U\left(\xi-y\right) + \lambda\right)}}.\label{eq:Saddle_point_phi_final_main}\\
   \mathrm{and} \ f&=\int dx\, \int{\cal D}t_{x}\,\frac{1}{1+e^{-\left(\phi_{x}+t_{x}g\sqrt{2q}+\U\left(\xi-x\right) + \lambda\right)}},\label{eq:Saddle_point_lambda_final_main}
\end{align}
with the Gaussian measure
\begin{align}
    \int{\cal D}t = \frac{1}{\sqrt{2\pi}}\int dt e^{-\frac{t^2}{2}}. 
\end{align}  
The entire statistics of our system can now be determined by taking derivatives of $G$ with respect to $h$, which is set to $0$ afterwards. We have to calculate the total derivative, also taking into account the $h$-dependence of $q$, $\phi$ and $\lambda$, which in turn, by the implicit-function theorem, we obtain by taking the total derivative with respect to $h$ of the their saddle-point-equations. This yields 
\begin{align}
    \label{eq:Covariances_incl_all_dependencies_main}
    \frac{1}{N}\frac{d^{2}}{d \boldsymbol{h}^{2}} \left\langle W_{f}\left(\boldsymbol{h}\right)\right\rangle_{J} = \frac{\partial^{2}G}{\partial\boldsymbol{h}^{2}}-\begin{pmatrix}\frac{\partial^{2}G}{\partial\boldsymbol{h}\partial\boldsymbol{\phi}}\\
\frac{\partial^{2}G}{\partial\boldsymbol{h}\partial q}\\
\frac{\partial^{2}G}{\partial\boldsymbol{h}\partial\lambda}
\end{pmatrix}^{\mathrm{T}}\begin{pmatrix}\frac{\partial^{2}G}{\partial\boldsymbol{\phi}^{2}} & \frac{\partial^{2}G}{\partial\boldsymbol{\phi}\partial q} & \frac{\partial^{2}G}{\partial\boldsymbol{\phi}\partial\lambda}\\
\frac{\partial^{2}G}{\partial q\partial\boldsymbol{\phi}} & \frac{\partial^{2}G}{\partial q^{2}} & \frac{\partial^{2}G}{\partial q\partial\lambda}\\
\frac{\partial^{2}G}{\partial\lambda\partial\boldsymbol{\phi}} & \frac{\partial^{2}G}{\partial\lambda\partial q} & \frac{\partial^{2}G}{\partial\lambda^{2}}
\end{pmatrix}^{-1}\begin{pmatrix}\frac{\partial^{2}G}{\partial\boldsymbol{\phi}\partial\boldsymbol{h}}\\
\frac{\partial^{2}G}{\partial q\partial\boldsymbol{h}}\\
\frac{\partial^{2}G}{\partial\lambda\partial\boldsymbol{h}}
\end{pmatrix}.
\end{align}
Evaluating this expression, we obtain
\begin{align}
	C &=  \Kv + \Kv\Jspace_{\mathrm{eff}}\Kv + C^{\mathrm{indirect}}\label{eq:Covariance_contributions}\\
	& = \Kv \left(\mathbb{1} - K\Kv\right)^{-1} + C^{\mathrm{indirect}},
\end{align}
where $\Kv$ is the diagonal matrix with the disorder-averaged single-neuron variances
\begin{align}
    \Kv_{xy} &= \delta_{xy} \kv_{x},\\
    \mathrm{where} \ \kv_{x} &\coloneqq \int {\cal D} t\frac{\partial m_{x}}{\partial \phi_{x}} = \int{\cal D}t\, m_{x} \left(1- m_{x}\right),
\end{align}
$m_{x}$ is the magnetization conditioned on the Gaussian helping variable $t$,
\begin{align}
	m_{x} \coloneqq \frac{1}{1+e^{-\left[\phi_{x} + t\, g\sqrt{2q} + \U\left(\xi - x\right) + \lambda \right]}},
    \label{eq:Def_mx}
\end{align}
the effective local connectivity $\Jspace_{\mathrm{eff}}$ is given by $\left[\left(\Jspace^{\mathrm{eff}}\right)^{-1}\right]_{xy}  \coloneqq -\frac{\partial^{2}G}{\partial\phi_{x}\partial\phi_{y}}$ and fulfills the Dyson equation
\begin{align}
	\Jspace_{xy}^{\mathrm{eff}} & =\Jspace_{xy} + \int dz\,  \Jspace_{xz} \kv_{z} \Jspace_{zy}^{\mathrm{eff}},
\end{align}
and $C^{\mathrm{indirect}}$ emerges from the remaining part of the Hessian in eq. (\ref{eq:Covariances_incl_all_dependencies_main}). It results in an subleading contribution to the Fisher information (see below), so we give its precise form only in appendix \ref{sec:CompFisherInfo}.
\subsection{Disorder-averaged Fisher information}
The Fisher information per neuron averaged over the disorder now reads
\begin{align}
	{\cal I}_{\boldsymbol{n}}\left(\xi\right) =  \sum_{x} \, \sum_{y} \, \U^{\prime} \left(\xi - x\right) \left[\kv_{x}\delta_{x,y} + \kv_{x} K^{\mathrm{eff}}_{xy} \kv_{y}  + C^{\mathrm{indirect}}_{xy}\right]  \U^{\prime}\left(\xi - y\right)
    \label{eq:FisherInfo_main}
\end{align}
In fig. (\ref{fig:FisherInfo_vs_g_theo_different_contributions}) in the appendix, we show these three contributions separately for the parameters used for fig. (\ref{fig:FisherInfo_vs_g_MC_comparison}). The first term stems from the single-neuron variances and is therefore also present without network (if present the variances are effected by the network, though). The third is always negligible, which intuitively makes sense because it emerges from the indirect $\boldsymbol{h}$-dependence of the free energy via $g$ and $\lambda$, which are both spatially unstructured. The second term emerges from the (positive) local interactions and also contributes positively. 
In order to gain a better intuition for where this term comes from, it is useful to re-derive the expression for the Fisher information using eq. (\ref{eq:FisherInfo_with_tuning_curves}), limiting ourselves to the case without disorder. By some lines of rearrangements, shown in appendix \ref{sec:Derivation_input_tuning_curve_Keff}, we derive that the vector of the derivatives of the tuning curves can be expressed as 
\begin{equation}
    \boldsymbol{\Tc}^{\prime}=V\left(1+K_{\mathrm{eff}}V\right)\boldsymbol{\U}^{\prime}.
    \label{eq:Tuning_curve_with_Keff}
\end{equation}
Combining this expression with eq. (\ref{eq:FisherInfo_with_tuning_curves}) and $C = V + V K_{\mathrm{eff}} V$, we are getting back eq. (\ref{eq:FisherInfo_main}) (without the contribution of the disorder) after canceling a factor $\left(V + VK_{\mathrm{eff}} V\right)$, as expected. However, it is also insightful to write down the expression before this cancellation,
\begin{equation}
    {\cal I}_{\boldsymbol{n}} = \overbrace{\left(\boldsymbol{\U}^{\prime}\right)^{\mathrm{T}}\left(V+VK_{\mathrm{eff}}V\right)}^{=\left(\boldsymbol{\Tc}^{\prime}\right)^{\mathrm{T}}}\overbrace{\left[V+VK_{\mathrm{eff}}V\right]^{-1}}^{=C^{-1}}
    \overbrace{\left(V+VK_{\mathrm{eff}}V\right)\boldsymbol{\U}^{\prime}}^{=\boldsymbol{\Tc}^{\prime}},
    \label{eq:FisherInfo_uncancelled}
\end{equation}
because it gives an intuition about how the local connectivity shapes the Fisher information: first, it modifies the tuning curves, which is captured by the term $V K_{\mathrm{eff}} V \U^{\prime}$, second it introduces cross-covariances, which is captured by the term $V K_{\mathrm{eff}} V$ contributing to the covariance. 
\begin{figure}
	\centering
    \subfloat{{\includegraphics[width=.5\textwidth]{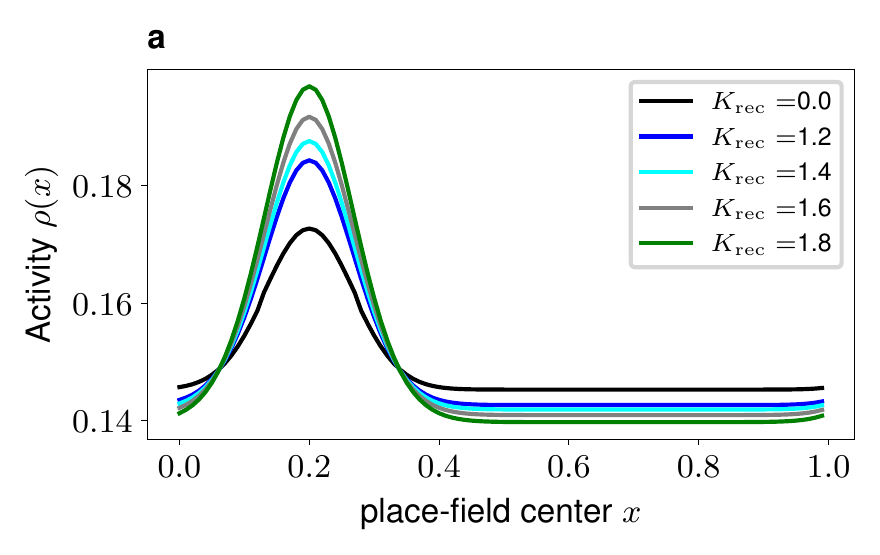} }}
    \subfloat{{\includegraphics[width=.5\textwidth]{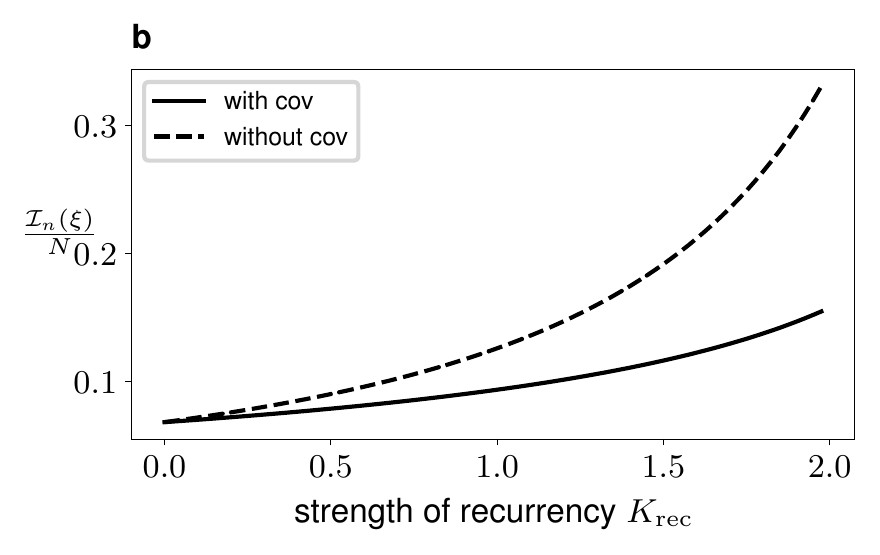} }}
     \caption{Panel a: tuning curves for a network without disorder in the connectivity for different strengths of local interactions. Panel b: corresponding change in the Fisher information; the dotted lines show eq. (\ref{eq:FisherInfo_uncancelled}) with the term $VK_{\mathrm{eff}}V$ in the middle part, constituting $C^{-1}$, removed. Parameters: $g=0$, $w_{\mathrm{two \ point}} = 0.1$, $\Aonepoint = 0.2$, $w_{\mathrm{one \ point}} = 0.07$, $f=0.15$.}
    \label{fig:Scale_local_conn_without_disorder}
\end{figure}
As apparent from fig. (\ref{fig:Scale_local_conn_without_disorder}), panel a, the tuning curves are sharpened with increasing $\Atwopoint$, which is reflected by the fact that the direct contribution of the cross-covariances to the Fisher information is positive, see fig. (\ref{fig:FisherInfo_vs_g_theo_different_contributions}). The cross cavariances, in turn, are detrimental in our case: they reduce the Fisher information, as apparent from panel b in fig. (\ref{fig:Scale_local_conn_without_disorder}).
\\
We can study further eq. (\ref{eq:FisherInfo_main}) analytically, which amounts to examining the saddle-point equations (\ref{eq:Saddle_point_q_main}) to (\ref{eq:Saddle_point_lambda_final_main}). In particular, we can do this in the limiting case $g \rightarrow 0$. In this limit, the Gaussian integrals get trivial. As detailed in appendix \ref{sec:Analysis_SP_eq}, we can use for the study of the derivatives of $q$, $\phi$ and $\lambda$ that, in eqs. (\ref{eq:Saddle_point_q_main}) to (\ref{eq:Saddle_point_lambda_final_main}), these quantities are (implicitly) given by
\begin{align}
    0 =  \frac{\partial}{\partial\left[\left\{\phi_{x}\right\}_{x},\lambda,q\right]} G_{g} \left(g, q, \phi, \lambda\right),
\end{align}
with $G$ as given in eq. (\ref{eq:Cumulant_generating_functional_thermodynamic_limit}) and therefore, by the implicit-function theorem,
\begin{align}
    \frac{\partial}{\partial g} 
	\begin{pmatrix}
    		\left\{\phi_{x}\right\}_{x}\\
    		\lambda\\
    		q
    	\end{pmatrix}
	= - \left(\frac{\partial^2}{\partial \left[\left\{\phi_{x}\right\}_{x},\lambda,q\right]^{2}} G_{g}\left[q,\boldsymbol{\phi},\lambda\right]\right)^{-1}
	\frac{\partial^{2}}{\partial g \partial \left[\left\{\phi_{x}\right\}_{x},\lambda,q\right]}G_{g}\left[q,\boldsymbol{\phi},\lambda\right].
\end{align}
For a reasonable choice of the parameters, also guaranteeing the stability of the saddle-point solution, $\frac{\partial^2}{\partial \left[\left\{\phi_{x}\right\}_{x},\lambda,q\right]^{2}} G_{g}$ is invertible (which we as well check numerically by computing it explicitly in appendix \ref{sec:CompFisherInfo}). The partial derivative of $\partial_{\left\{\phi_{x}\right\}_{x},\lambda,q}G$ with respect to $g$ vanishes in the limit $g\rightarrow 0$, as we show in appendix \ref{sec:Analysis_SP_eq}. Therefore, the derivatives of $q$, $\phi$ and $\lambda$ with respect to $g$ go to $0$ for vanishing $g$. This results carries over to higher-order cumulants and to the cumulant-generating functional itself. Because the Fisher information depends on $g$ only via these quantities, its derivative vanishes in the limit of $g \rightarrow 0$:
\begin{equation}
	\underset{g \rightarrow 0}{\lim}\, \frac{\partial}{\partial g}{\cal I}_{\boldsymbol{n}}\left(\xi\right)=0.
	\label{eq:Limit_FisherInfo_deriv_g_g_to_0}
\end{equation}
 The derivatives with respect to the strength of the local interaction, $\Atwopoint$, however, in general do not vanish for vanishing connectivity. Therefore even small connection strengths will have a (beneficial) effect, as seen before.

\section{Numerical validation of mean-field results and applications \label{sec:Validation_Application}}
\subsection{Monte Carlo simulation}
\begin{figure}
	\centering
    \subfloat{{\includegraphics[width=.5\textwidth]{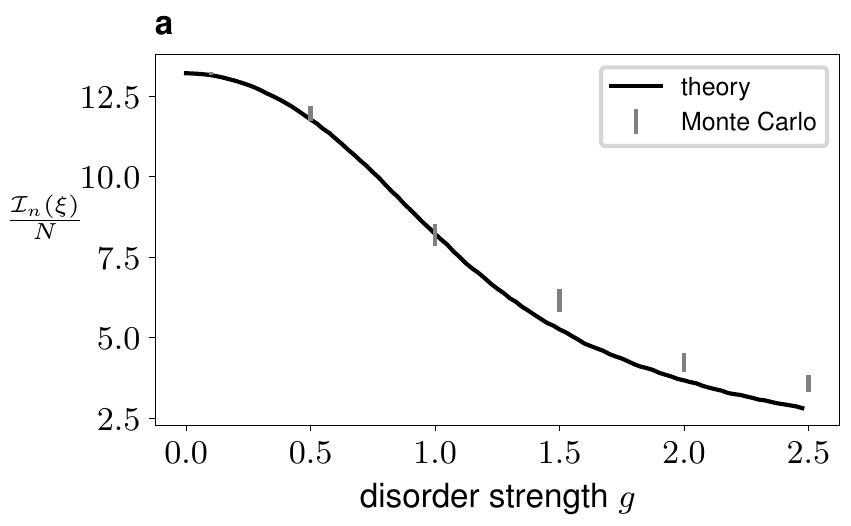} }}
    \subfloat{{\includegraphics[width=.5\textwidth]{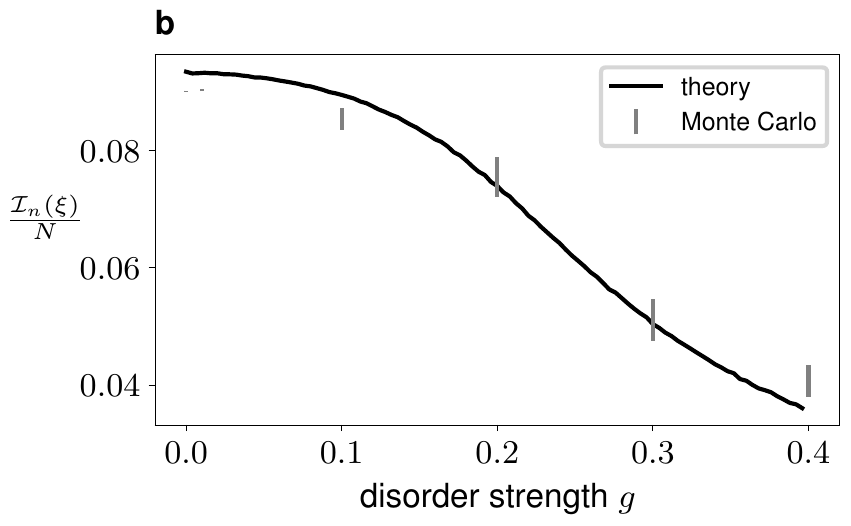} }}
     \caption{The Fisher information per neuron in dependence of the disorder, comparison with Monte-Carlo simulations. In panel a, the network is strongly input driven, and in panel b only weakly. 
     Parameters, panel a: $\Atwopoint = 5$, $\wtwopoint = 0.1$, $\Aonepoint = 2.25$, $\wonepoint = 0.07$, $f=0.15$; 
     panel b: $\Atwopoint = 20$, $\Aonepoint = 0.2$, other parameters as in panel a.}
	\label{fig:FisherInfo_vs_g_MC_comparison}
\end{figure}

To validate our computation derived for the thermodynamic limit, we perform Monte-Carlo simulations for multiple sets of parameters, see fig. (\ref{fig:FisherInfo_vs_g_MC_comparison}).
   We use a standard Metropolis algorithm, taking into account the condition of a fixed total activity by flipping always two spins, in opposite direction, as suggested in \cite{Rosay13_062813}. The code for these simulations, alongside with the implementation of the mean-field results is provided in \cite{Kuehn23_SpacentationCode}. The results confirm our mean-field result that the disorder diminishes the Fisher information, but quite slowly if the disorder is on a moderate level, as predicted by eq. (\ref{eq:Limit_FisherInfo_deriv_g_g_to_0}) and visible in  fig. (\ref{fig:FisherInfo_vs_g_MC_comparison}).

\subsection{Influence of the network on the Fisher information}
\begin{figure}
	\centering
    \subfloat{{\includegraphics[width=.5\textwidth]{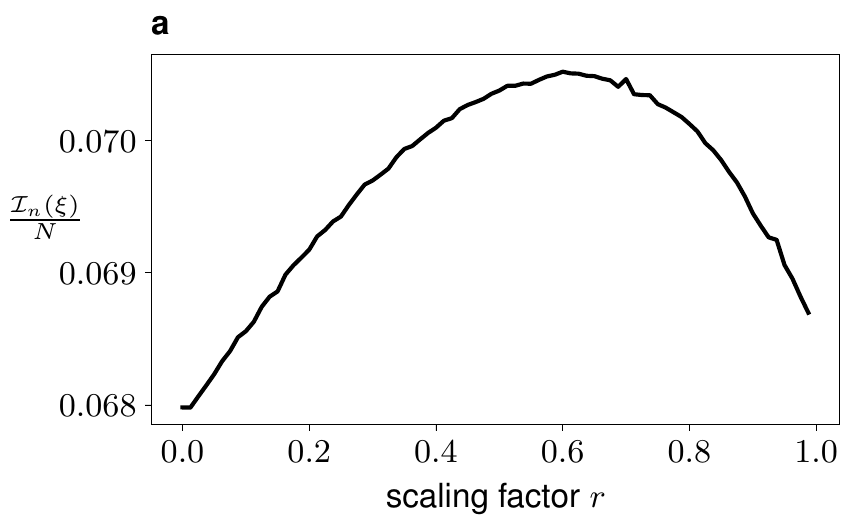} }}
    \subfloat{{\includegraphics[width=.5\textwidth]{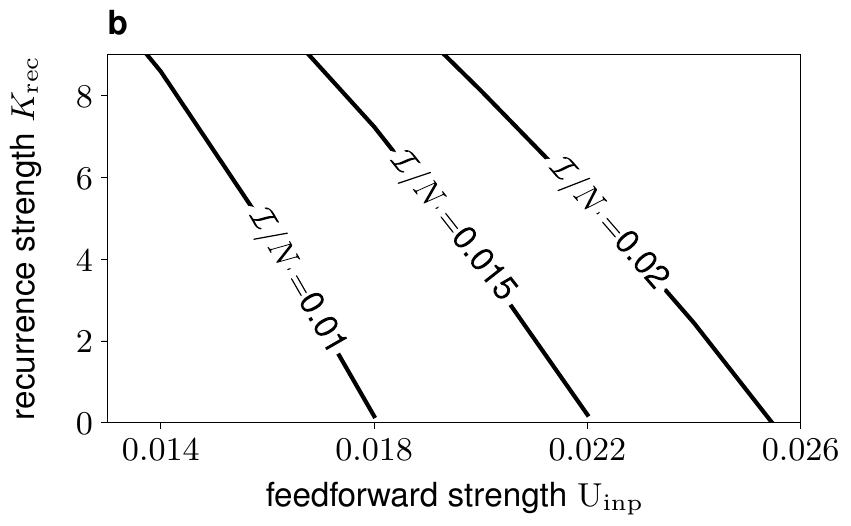} }}
     \caption{Interplay of feedforward, local and disordered recurrent input shaping the  Fisher information. For panel a, we scale the synapses according to eq. (\ref{eq:scale_synapses}), keeping the other parameters fixed, for panel b, we keep the the Fisher information constant, varying $\Aonepoint$ and $\Atwopoint$ concertedly.
     Parameters, panel a: $\Atwopoint^{\mathrm{max}} = 8$, $\wtwopoint = 0.1$, $\Aonepoint = 0.2$, $\wonepoint = 0.07$, $f=0.15$, $g^{\mathrm{max}} = 0.16$
     panel b: $g = 5$, other parameters as in panel a.}
	\label{fig:Fisher_Info_synaptic_scaling_and_isoFisherInfo}
\end{figure}

In an attractor network, disorder may result from the presence of other maps stored in the same network. Therefore it scales in the same way as the spatially dependent part of the connectivity. It is thus interesting to examine the behavior of the Fisher information when both parts of the connectivity are scaled by the same factor $r$:
\begin{equation}
	\Jspace \rightarrow r\cdot \Jspace, \quad \Jdisord \rightarrow r\cdot \Jdisord.
	\label{eq:scale_synapses}
\end{equation}
Because the derivative of the expression for the Fisher information with respect to the disorder strength $g$ vanishes for $g=0$, the effect of the local part of the connectivity dominates for small synaptic strength: the Fisher information initially increases. This can be understood from what we have derived before: increasing the local connectivity sharpens the tuning curves and therefore increases the signal. This effect is diminished (but not cancelled) by the introduction of covariances between the neurons (see also \cite{Pouget99_85}). For larger scaling factors, however, this overall beneficial effect is wiped out by the disorder, whose detrimental effect eventually dominates, fig. (\ref{fig:Fisher_Info_synaptic_scaling_and_isoFisherInfo}), panel a.  

Finally, we ask if we can keep the Fisher information constant by increasing the recurrent weights when the input gets weaker. We have plotted lines of constant Fisher information for varying strength of the input and the recurrent connections in fig. (\ref{fig:Fisher_Info_synaptic_scaling_and_isoFisherInfo}), panel b. We can indeed make up to a decrease in the input by strengthening local connections, even though of course only in a limited range.

\subsection{\label{sec:Linread} Linear readout}
\begin{figure}
	\includegraphics[width=.5\textwidth]{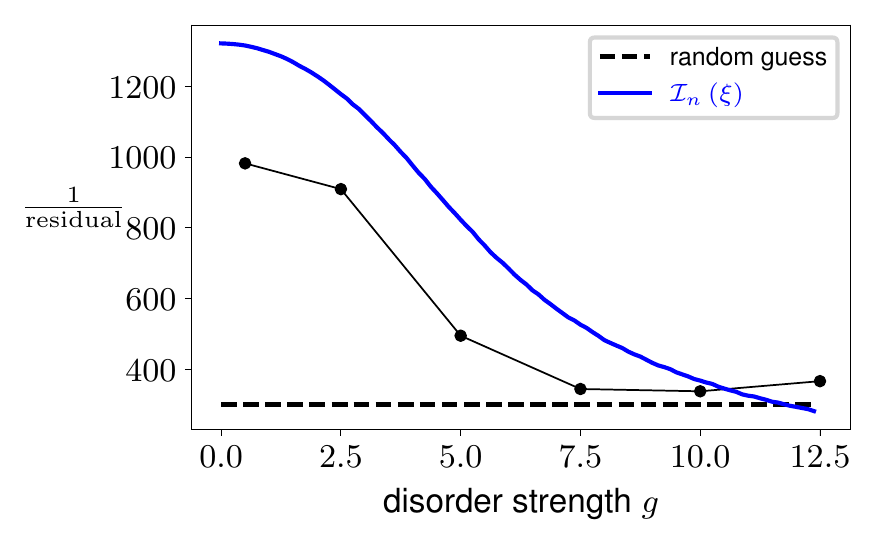}
	\caption{Inverse residual of a linear fit of the neural activity to estimate $\xi$, averaged over $\xi \in \left[0.4, 0.6\right]$. Only for small $g$ (approximately the first two data points at $0.1$ and $0.5$), this estimate is approximately unbiased and therefore only there, the Cram\'er-Rao bound guaranties that the Fisher information is an upper bound. Note that in this plot, different to all the others, we are plotting the total Fisher information for all $100$ neurons, not the averaged, single-neuron analog. Parameters as in fig. \ref{fig:FisherInfo_vs_g_MC_comparison}, panel a.}
	\label{fig:Inv_linear_readout_error_vs_FisherInfo}
\end{figure}
To put our results into context and convey a more intuitive understanding, we briefly discuss what one can learn from the Fisher information about the accuracy of a linear readout.

We have fitted a readout vector $\ro$ to the activity as measured after the thermalization process (so the random initial conditions should not play a role) and computed the squared residual error as 
\begin{align}
	\left\langle\mathrm{res}\left(\xi\right)\right\rangle_{\xi} \coloneqq \underset{\ro}{\min} \left \langle \left \langle \left \|\ro\cdot \boldsymbol{n} - \xi \right \|_{2}^{2} \right\rangle_{\boldsymbol{n}}\right \rangle_{\xi} 
	 &=  \left \langle \xi^{2}\right\rangle_{\xi} -2\ro_{\mathrm{train}}^{\mathrm{T}}C^{\mathrm{test}}_{\xi,\boldsymbol{n}} +  \ro_{\mathrm{train}}^{\mathrm{T}}C^{\text{test}}_{\boldsymbol{n},\boldsymbol{n}}\ro_{\mathrm{train}} \\
	\mathrm{where} \ \ro_{\mathrm{train}} & = \left[C^{\mathrm{train}}_{\boldsymbol{n},\boldsymbol{n}} + \lambda \cdot  \mathbb{1}\right]^{-1} C^{\mathrm{train}}_{\xi,\boldsymbol{n}},   
\end{align}
where $\left \langle \dots\right\rangle_{\boldsymbol{n}}$ and $\left \langle \dots\right\rangle_{\xi}$ denote the thermal average over the configurations $\boldsymbol{n}$ and the average over the distribution of the stimulus $\xi$, respectively and $\lambda$ denotes the strength of the $L_{2}$ regularization that we impose. We might expect to obtain an upper bound for the accuracy of the linear estimator by the Cram\'er-Rao bound. However, due to the periodic boundary conditions, the estimate from the linear readout $\xi_{\mathrm{est.}} = \ro\cdot \boldsymbol{n} $ will be biased. This is particularly apparent at the borders $0$ and $1$, where the estimate will always be $\frac{1}{2}$, corresponding to random guessing. The farther away from the them the stimulus is situated, the less pronounced the effect becomes. We have therefore limited the fitted stimuli to $\xi \in \left[0.4, 0.6\right]$. However, even in this regime, the linear readout gets biased in the highly disordered regime, so that there the Cram\'er-Rao bound only applies in its generalized form \citep{Cover91}, their eq. (12.333):
\begin{align}
	\left\langle \left(\ro\cdot \boldsymbol{n} - \xi\right)^{2} \right \rangle_{\boldsymbol{n}} \geq \frac{\left(1+b^{\prime}\left(\xi\right)\right)^{2}}{{\cal I}_{\boldsymbol{n}}\left(\xi\right)}+b\left(\xi\right)^{2},
\end{align}
where $b\left(\xi\right) \coloneqq \left\langle \ro\cdot \boldsymbol{n}\right\rangle_{\xi} - \xi $ is the bias of the linear estimator. In case of random guessing, in particular, $b^{\prime}\left(\xi\right)=-1$, so that the error is solely determined by the square of the bias. Therefore, the bound given by the Fisher information in fig. (\ref{fig:Inv_linear_readout_error_vs_FisherInfo})
 is only meaningful for small disorder ($g\sim 0.1, 0.5$), whereas for greater disorder, it is invalidated by the bias, up to the point where the linear readout basically generates a random guess (for $g \sim 2$). In the low-disorder regime, however, we observe that the error of the linear readout is not far from the optimal case given by the Cram\'er-Rao bound.

\section{\label{sec:Discussion} Discussion and Outlook}

In this work, we have studied an attractor-inspired neural network with a connectivity consisting of two parts: (1) excitatory couplings between neurons that are similarly tuned to stimuli, and (2) a quenched random background without local tuning. We have studied the influence of both of these contributions on the information about the stimulus contained in the neural activity through the analytical computation of the Fisher information. As expected the local part of the connectivity enhances the information content, whereas the disordered part degrades it. However, the latter effect is mild. By fitting a linear readout to estimate the driving stimulus we show that the Fisher information is not only a formal estimate of the information contained in the neural activity, but also gives a useful bound on how much of it can be extracted with simple decoders.

It has been recognized for a long time that the presence of disorder in the interactions could impact the information stored in attractor networks in the form of patterns. In the case of CANNs this translates into a breaking of the translational symmetry of bump-like solutions, see for instance \cite{Tsodyks95} and box 3 in \cite{Moser14_466}. However, despite the loss of translation invariance, noise in the neural activity (controlled here by the inverse of the coupling strength, playing the role of temperature in statistical mechanics) may be sufficient to move the bump \cite{Monasson14_032803}. We here show that the disorder does not wipe out all information in the attractor network. In particular, the Fisher information is robust to the introduction of disorder, staying constant to first order in $g$. Consequently, globally enhancing the connectivity strength (the local and the disordered part by the same factor, as in eq. (\ref{eq:scale_synapses})) initially always has a beneficial effect, which is overtaken by the effect of the disorder only for larger connectivity. In real biological networks, the connectivity is not fixed, but builds up during development,  partially through learning \cite{Richter17_39}. One might therefore speculate if this process is optimized for the synaptic strengths to eventually match this sweet spot.

Similarly, we can ask for the information-theoretical implication of the development of tuning curves as observed in the visual cortex. In young animals, recurrent connections between similarly tuned neurons get enhanced, other weakened \cite{Ko13_96}, and the orientation selectivity is sharpened \cite{Sadeh15_1}. According to our analysis, these changes in recurrent connectivity could compensate a decrease in strength of sensory inputs  \cite{Lim19_e44098}, see fig.~\eqref{fig:Fisher_Info_synaptic_scaling_and_isoFisherInfo}(b).

While our model allows for this kind of qualitative considerations, it is minimal in the sense that it contains the ingredients needed to study the effects we are interested in in their simplest form. This of course limits  biological plausibility and calls for enhancements. From a technical point of view, we did of course not consider all possible scenarios. We give below an outlook on possible further directions, starting with the technical aspects. 

The results we have obtained here hold for binary-valued neurons. From a neuroscience perspective, this assumption can be interpreted as follows. Consider the neural population activity in a time-bin of duration $\Delta t$. If $\Delta t$ is small enough (in practice, not larger than the inverse of the typical firing rate), it is likely that each neuron $i$ has fired at most once in a time bin. We can therefore represent its activity through a binary variable $n_i$ equal to $0$ in the absence of spike, or to $1$ if a spike has been emitted. This point of view was adopted in data-driven models of the hippocampal activity \cite{posani18}. If the duration $\Delta t$ is large, the binary hypothesis breaks down and continuous models, taking into account the real-value nature of firing rates, should be considered. Continuous-attractor models with real-valued neurons show similar - albeit for some parameters a bit richer \cite{Treves90_2631} - behavior as models with binary neurons \cite{Tsodyks95}, so we do not expect our conclusions to be qualitatively modified in that setting. In addition, binary patterns represent an important limiting case: they may maximize the  retrievable information, in the case of random memories, as compared to more complex ones \cite{Treves90_2418}. 

Then, we have limited ourselves to a parameter regime far away from phase transitions. In particular we have not considered  the low-temperature/high-connectivity regime. The analysis of this case (without the computation of the Fisher information) was carried out by one of us and collaborators in a series of  papers \cite{Rosay13_062813, Monasson14_032803, Monasson15_098101}, in the specific case of background noise due to a extensive number $\alpha N$ of alternative attractors embedded in the network. Although the quenched noise distribution in this case was not Gaussian we do not expect the phase diagram to significantly change.  Based on these previous works we can thus build educated guesses on what to expect in terms of information theory. Note that, in terms of scaling, the square of the disorder strength used here, $g^{2}$, roughly corresponds to the load $\alpha$ in \cite{Rosay13_062813} (compare, e.g. their saddle-point equations, eq. (28) to ours, eqs. (\ref{eq:Saddle_point_q_main}) to (\ref{eq:Saddle_point_lambda_final_main})).

Analog to \cite{Rosay13_062813}, we expect a glassy phase for large disorder and weak local connectivity, a ferromagnetic ("bump") phase for weak disorder and strong local connectivity and a paramagnetic phase in case both contributions are small (large-temperature regime). In the present study, we have basically stayed in the latter regime (however, the activity was still bump-like due to the feed-forward input).
This limitation has also allowed us to stick to the replica-symmetric solution of the saddle-point equations - an assumption that might  not be satisfied at very low temperature and strong enough disorder; however, the comparison with the results of Monte-Carlo simulations in fig. (\ref{fig:FisherInfo_vs_g_MC_comparison}) led to reasonable results. This is expected; for spin-glass models, the effect of replica-symmetry breaking is  typically rather limited, in particular close to the Almeida-Thouless line. In order to study the glassy regime, corresponding to a network load $\alpha$ beyond the critical value, we expect that replica symmetry breaking has to be considered. However, as for the Fisher information, we expect  it to be approximately zero in the glassy state because the activity would not show any dependence on space anymore (besides a residual one due to the input), and the disorder-averaged single-neuron variances vanish for large disorder (and all other cumulants as well). As for the paramagnetic to ferromagnetic transition, we expect a qualitative change in the shape of the neural activities, with considerably sharper bumps on the ferromagnetic side of the transition. Due to the mechanisms discussed around eq. (\ref{eq:FisherInfo_with_tuning_curves}), we expect a corresponding steep increase in the Fisher information.

On a more biological side, we have made, for convenience, several unrealistic assumptions that could be waived. The receptive fields in our setup, for example, have all the same shape and size and are evenly spaced - in reality there is of course variability in  shape and they are scattered in the environment. In particular, place fields may cluster near new objects  \cite{Bourboulou19_e44487}, suggesting the importance of taking into account  inhomogeneous densities. These features could rather easily be included in our framework, at least if the probability distributions for the single-neuron properties are independent. The additions suggested above only require another average over them (see also appendix \ref{sec:MutualInfo_FisherInfo_independent}). Then, though we have here studied a one-dimensional stimulus, the features determining the Fisher information, such as the sharpness of the tuning curve, can also be defined in higher dimensions, and we expect qualitatively the same results in that case. Last of all, it would be interesting to better understand  how much of the information we have estimated can be extracted from the neural population in practice, beyond the linear readout mechanism considered here. 

\begin{acknowledgments}
TK thanks Ulisse Ferrari and Gabriel Mahuas for many very insightful discussions. This work was partly funded by the Human Frontier
Science Program RGP0057/2016 grant and TK by a short-term postdoc fellowship of the German Academic Exchange Service (DAAD).
\end{acknowledgments}

\appendix

\section{Computing the cumulant-generating function}
As indicated in the main text, computing the Fisher information basically amounts to calculating the covariance matrix of the activity, which we obtain from the disorder-average of the cumulant-generating functional defined in eq. (\ref{eq:Def_disorder_averaged_cumulant_gen_fct}), or in other words, of the logarithm of the partition function. In the following, we will explain step by step how to take into account its features, starting with the fixed total activity, then including the disordered part of the connectivity and finally deriving and analysing the saddle-point equations for this case in order to compute the cumulant-generating functional in the thermodynamic limit. 
\subsection{Effects of the fixed total activity and the space-dependent part of the coupling}
\label{sec:App_fixedAct_spacedep_coupl}
As mentioned in the main text, we fix the summed activity of all neurons to a certain number $M$,
mimicking the effect of a global inhibition. This determines the
partition function, which reads
\[
{\cal Z}_{\Nact}\left[\boldsymbol{h}\right]=\sum_{\boldsymbol{n},\sum_{i}n_{i}=\Nact}\exp\left(\sum_{i<j,}\Jspace_{ij}n_{i}n_{j}+\sum_{i=1}^{N}h_{i}n_{i}\right).
\]
Explicitly performing the spin sums under this limitation is difficult, so we introduce the Fourier series with ${\cal Z}_{\Nact}\left[\boldsymbol{j}\right]$ as coefficients:
\begin{align*}
U_{k}\left[\boldsymbol{h}\right] & :=\sum_{\Nact=0}^{N}\,e^{i2\pi k\Nact}{\cal Z}_{\Nact}\left[\boldsymbol{h}\right]\\
 & =\sum_{\Nact=0}^{N}\,e^{i2\pi k\Nact}\sum_{\boldsymbol{n}}\delta_{\Nact,\sum_{i}n_{i}}\exp\left(\sum_{i<j,} \Jspace_{ij}n_{i}n_{j}+\sum_{i=1}^{N} h_{i}n_{i}\right)\\
 & =\sum_{\boldsymbol{n}}\exp\left(\sum_{i<j,}\Jspace_{ij}n_{i}n_{j}+\sum_{i=1}^{N} h_{i} n_{i}\right)\sum_{\Nact=0}^{N}\,e^{i2\pi k\Nact}\delta_{\Nact,\sum_{i}n_{i}}\\
 & =\sum_{\boldsymbol{n}}\exp\left(\sum_{i<j,}\Jspace_{ij}n_{i}n_{j}+\sum_{i=1}^{N}\left(h_{i}+i2\pi k\right)n_{i}\right).
\end{align*}
Applying the transform to obtain the Fourier coefficients from a periodic function, we get
\begin{align*}
{\cal Z}_{\Nact}\left[\boldsymbol{h}\right] & = \int_{0}^{1}dk\, e^{-i2\pi k\Nact}U_{k}\left[\boldsymbol{j}\right]\\
 & = \int_{0}^{1}dk\,\sum_{\boldsymbol{s}}\exp\left(\sum_{i<j,}\Jspace_{ij}n_{i}n_{j}-i2\pi k\left(\Nact-\sum_{i=1}^{N}n_{i}\right)+\sum_{i=1}^{N}h_{i}n_{i}\right).
\end{align*}
For $N\gg1$, we can evaluate the $k$-integral in saddle-point approximation,
so that in this limit, we replace the partition function by its
``grandcanonical'' counterpart
\begin{align*}
{\cal Z}_{f,\mathrm{gc}}\left[\boldsymbol{h}\right] & :=\inf_{\lambda}\left[\sum_{\boldsymbol{n}}\exp\left(\sum_{i<j,}\Jspace_{ij}n_{i}n_{j}+\sum_{i=1}^{N}h_{i}n_{i}+\lambda\left(\sum_{i=1}^{N}n_{i}-Nf\right)\right)\right],
\end{align*}
where we have introduced $f:=\frac{\Nact}{N}$ and $\lambda = i2\pi k$. Note that even though the stationary value of $\lambda$ is real, we are  integrating it along the imaginary axis (using what is known as Bromwich contour, see e.g. \cite{Touchette09_1}, appendix C), as always when the integration variable has been introduced as Lagrance multiplyer. Varying $\lambda$, we are therefore looking for the infimum, not the supremum. We obtain the ``grand-canonical'' cumlant-generating function that we will work with:
\begin{align*}
W_{\lambda}\left[\boldsymbol{h}\right] & :=\ln\left[\sum_{\boldsymbol{s}}\exp\left(\sum_{i<j,}\Jspace_{ij}n_{i}n_{j}+\sum_{i=1}^{N}\left[h_{i}n_{i} + \lambda \left(n_{i} - f\right)\right]\right)\right]\\
\lambda_{\boldsymbol{h},\boldsymbol{j},\Jspace}\left(f\right)\mathrm{\ such\ that\ \frac{\partial W_{\lambda}\left[\boldsymbol{h}\right]}{\partial\lambda}} & =Nf.
\end{align*}
Now we decouple the interacting term by means of a Gaussian helping
field
\begin{align*}
 & \exp\left(\frac{1}{2}\sum_{i\neq j,}\Jspace_{ij}n_{i}n_{j}\right)\\
= & \frac{1}{\left(2\pi\right)^{\frac{N}{2}}\sqrt{\det\left(\Jspace\right)}}\int d\boldsymbol{\phi}\,e^{-\frac{1}{2}\boldsymbol{\phi}^{\mathrm{T}}\Jspace^{-1}\boldsymbol{\phi}+\sum_{i}\phi_{i}n_{i}}
\end{align*}
and replace the $\phi$-integral by another saddle-point approximation:
\begin{align*}
W\left[\boldsymbol{h}\right] & =\sup_{\boldsymbol{\phi}}\, \inf_{\lambda}\left[-\frac{1}{2}\boldsymbol{\phi}^{\mathrm{T}}\Jspace^{-1}\boldsymbol{\phi}+\sum_{i}\left(\ln\left(1+e^{h_{i}+\lambda+\phi_{i}}\right)-\lambda f\right)\right]
\end{align*}
For the examples shown in the figures, we are assuming a rectangular shape for $K$,
\begin{align}
	\Jspace \left(r_{i} - r_{j}\right) = \begin{cases}
		\Atwopoint, & \mathrm{for} \ \left| r_{i} - r_{j} \right| \leq \wtwopoint \\
		0, & \mathrm{else},
	\end{cases}\label{eq:Def_determin_connect}
\end{align}
but this choice is only made for convenience, the theoretical results extend to general shapes.
\subsection{\label{app:disorder_saddle_point}Incorporating disorder}
Drawing random connections in addition to the spatially ordered ones additionally modifies the extremizing probability distribution and introduces more contributions to the pairwise covariances. We would like to compute the quenched average of the cumulant-generating functional, so
\begin{align}
    \left\langle W\left(\boldsymbol{h}\right)\right\rangle_{J}= & \int dJ\,P\left(J\right)\ln\left[\sum_{\boldsymbol{n}}e^{\frac{1}{2}\sum_{i\neq j}\left(\Jdisord_{ij}+\Jspace_{ij}\right)n_{i}n_{j}+\sum_{i}\left(n_{i}\U\left(\xi-r_{i}\right)+h_{i}n_{i}\right)}\right]\\
= & \lim_{n\rightarrow0} \int dJ\,P\left(J\right)\label{eq:Cumulant_gen_fct_disorder_average} \left[\frac{-1+\sum_{\boldsymbol{n}^{1},\dots,\boldsymbol{n}^{n}}e^{\sum_{i\neq j}\left(\Jdisord_{ij}+\Jspace_{ij}\right)\sum_{\alpha=1}^{n}n_{i}^{\alpha}n_{j}^{\alpha}+\sum_{i}n^{\alpha}_{i}\left(\U\left(\xi-r_{i}\right)+h_{i}\right)}}{n}\right],
\end{align}
where we have used the replica trick to represent the logarithm \cite{Mezard87}. As indicated in the main text, eq. (\ref{eq:statistics_disorder}), we assume that the couplings are uncorrelated and Gaussian, so that, after the standard procedure of introducing appropriate helping fields, we obtain
\begin{align}
    & \int dJ\,P\left(J\right)\,e^{\sum_{i\neq j}J_{ij}\sum_{\alpha=1}^{n}n_{i}^{\alpha}n_{j}^{\alpha}}\\
= & \exp\left(-\frac{1}{2}\frac{g^{2}}{N}\sum_{i}\sum_{\alpha,\beta}n_{i}^{\alpha}n_{i}^{\beta}\right)\prod_{\alpha}\left[\frac{\sqrt{N}}{g\sqrt{2\pi}}\int d\bar{q}_{\alpha}\exp\left(-\frac{1}{2}\frac{N}{g^{2}}\bar{q}_{\alpha}^{2}+\bar{q}_{\alpha}\sum_{i}n_{i}^{\alpha}\right)\right]\\
 & \times\prod_{\alpha\neq\beta}\left[\frac{\text{\ensuremath{\sqrt{N}}}}{g\sqrt{2\pi}}\int dq_{\alpha\beta}\,\exp\left(-\frac{1}{2}\frac{N}{g^{2}}q_{\alpha\beta}^{2}+q_{\alpha\beta}\sum_{i}n_{i}^{\alpha}n_{i}^{\beta}\right)\right].
\end{align}
We combine this result with the contribution from the network without disorder, but local connectivity and solve the resulting integral, assuming replica-symmetry in $q$ and $\boldsymbol{\phi}$. The validity of the assumption of replica symmetry is validated numerically by comparing our theoretical results with the outcomes of Monte-Carlo computations and discussed in sec. \ref{sec:Discussion}. Dropping subleading terms in $N$, we solve the integrals in saddle-point approximation and obtain for $W$:
\begin{align}
    \left\langle W\left(\boldsymbol{h}\right)\right\rangle _{J}= & \lim_{n\rightarrow0}\underset{q,\bar{q}, \bphi}{\mathrm{extr}}\left[-\frac{1}{n}
    +\frac{e^{-\frac{1}{2}Ng^{2}n\left(n-1\right)q^{2}
    -\frac{1}{2}Ng^{2}n\bar{q}^{2}}e^{-\frac{1}{2}n\boldsymbol{\phi}^{T}K^{-1}\boldsymbol{\phi}
    }}{n}\right.\\
 & \times\left.\left(\prod_{l,\gamma}\sum_{n_{l}^{\gamma}=0}^{1}\right)\left(\prod_{k}e^{g^{2}q\sum_{\alpha\neq\beta}n_{k}^{\alpha}n_{k}^{\beta}}\right)e^{\sum_{i}g^{2}\bar{q}\sum_{i}\sum_{\alpha}n_{i}^{\alpha} + \sum_{\alpha=1}^{n}n_{i}^{\alpha}\phi^{i}+\sum_{i}\sum_{\alpha=1}^{n}n_{i}^{\alpha}\left(\U\left(\xi-r_{i}\right)+h_{i}\right)}\right]\\
= & \frac{1}{2}Ng^{2}q^{2}-\frac{1}{2}Ng^{2}\bar{q}^{2}-\frac{1}{2}\boldsymbol{\phi}^{T}K^{-1}\boldsymbol{\phi}\\
 & +\prod_{k}\frac{1}{\sqrt{2\pi}}\int dt_{k}\,e^{-\frac{t_{k}^{2}}{2}} 
 \sum_{i}\ln\left[1+e^{\phi^{i}+t_{i}g\sqrt{2q}+\U\left(\xi-r_{i}\right)+g^{2}\left(\bar{q}-q\right) + h_{i}}\right]. 
\end{align}
Taking now into account the restriction on the total activity in addition, the mean-field cumulant-generating functional reads
\begin{align}
    \left\langle W\left(\boldsymbol{h}\right)\right\rangle_{J}
= & \underset{q,\bar{q}, \bphi, \lambda}{\mathrm{extr}} \left\{\frac{1}{2}Ng^{2}q^{2}-\frac{1}{2}Ng^{2}\bar{q}^{2}-\frac{1}{2}\boldsymbol{\phi}^{T}K^{-1}\boldsymbol{\phi}- N \lambda f\right.\\
 & \left.+\prod_{k}\frac{1}{\sqrt{2\pi}}\int dt_{k}\,e^{-\frac{t_{k}^{2}}{2}} \sum_{i}\ln\left[1+e^{\phi^{i}+t_{i}g\sqrt{2q}+\left(\U\left(\xi-r_{i}\right)\right) + g^{2}\left(\bar{q}-q\right)+ \lambda + h_{i}}\right]\right\}. 
\end{align}
Because we are evaluating this quantity only at its extremal values, we are free to express it in shifted coordinates, $\lambda + g^{2}\left(\bar{q} - q\right) \rightarrow \lambda$, in order to simplify our expressions and to get rid of $\bar{q}$, so that we obtain
\begin{align}
    \left\langle W_{I}\left(\boldsymbol{h}\right)\right\rangle_{J}
= & \underset{q,\bar{q}, \bpsi, \bphi, \lambda}{\mathrm{extr}} \left\{\frac{1}{2}Ng^{2}q^{2}-\frac{1}{2}Ng^{2}\bar{q}^{2}-\frac{1}{2}\bphi^{T}K^{-1}\bphi - N \left(\lambda - g^{2}\left(\bar{q} - q\right)\right) f.\right.\\
 & \left.+\prod_{k}\frac{1}{\sqrt{2\pi}}\int dt_{k}\,e^{-\frac{t_{k}^{2}}{2}} \sum_{i}\ln\left[1+e^{\phi^{i}+t_{i}g\sqrt{2q}+\U\left(\xi-r_{i}\right) + \lambda + h_{i}}\right]\right\}\\
 \eqqcolon & G_{g}\left(\boldsymbol{h}, \boldsymbol{\phi}, q, \lambda\right),\label{eq:Def_G_cum_gen_fct} 
\end{align} 
which leads to the saddle-point equations
\begin{align}
     q &= f + \int dx\, \int Dt\, \left\{\frac{1}{\left[1+e^{-\left(\phi_{x}+t_{x}g\sqrt{2q}+\U\left(\xi-x\right)+\lambda + h_{x}\right)}\right]^{2}} - \frac{1}{\left[1+e^{-\left(\phi_{x}+t_{x}g\sqrt{2q}+\U\left(\xi-x\right)+\lambda + h_{x}\right)}\right]}\right\}\label{eq:Saddle_point_q}\\
     \bar{q} &= f \label{eq:Saddle_point_qbar}\\
     f &= \int dx\,\int Dt\,\frac{1}{1+e^{-\left(\phi_{x}+t_{x}g\sqrt{2q}+\U\left(\xi-x\right)+\lambda + h_{x}\right)}}dx\label{eq:Saddle_point_lbda}\\
\phi\left(x\right) & =\int K\left(x-y\right)\int Dt\,\frac{1}{1+e^{-\left(\phi_{y}+t_{y}g\sqrt{2q}+\U\left(\xi-y\right)+\lambda + h_{y}\right)}}dy.\label{eq:Saddle_point_phi_final}
\end{align}
Drawing the limit of $N\rightarrow \infty$, we have turned the sums over neuron sites into integrals over space, which we indicate by renaming the indices to $x$ and $y$ instead of $i$ and $j$. Finally, setting $\boldsymbol{h}=0$ and using eq. (\ref{eq:Saddle_point_lbda}) to simplify eq. (\ref{eq:Saddle_point_q}), we obtain the final saddle-point equations as given in the main text, eqs. (\ref{eq:Saddle_point_q_main}) to (\ref{eq:Saddle_point_lambda_final_main}). Note that this simplification is valid for the saddle-point values $q$, $\left\{\phi_{x}\right\}_{x}$ and $\lambda$ - however, when taking further derivatives of $G$ with respect to $\boldsymbol{h}$ (as necessary to determine covariances), we have to assume general $q$, $\left\{\phi_{x}\right\}_{x}$ and $\lambda$ (not as given in the saddle point) and therefore have to use the right-hand side of eq. (\ref{eq:Saddle_point_q}), and not of eq. (\ref{eq:Saddle_point_q_main}).

\subsection{\label{sec:Analysis_SP_eq}Analysis of the saddle-point equations in the limit $g \rightarrow 0$}
The quantities $q$, $\phi_{x}$ and $\lambda$ are implicitly given by
\begin{equation}
	0 = \frac{\partial}{\partial \left[\left\{\phi_{x}\right\}_{x},\lambda,q\right]}G_{g}\left[q,\boldsymbol{\phi},\lambda\right].\label{eq:Implicit_def_auxiliary_var}
\end{equation}
For $g=0$, the integrands in the saddle-point equations (\ref{eq:Saddle_point_q_main}) to (\ref{eq:Saddle_point_lambda_final_main}) get independent of $t$ and we can perform the Gaussian integrals, so that we obtain
\begin{align}
     q & =\int dx\,\frac{1}{\left[1+e^{-\left(\phi_{x}+\U\left(\xi-x\right)t)+\lambda\right)}\right]^{2}},\label{eq:Saddle_point_q_g=0}\\
   \phi_{x} & =\int dy\, K\left(x-y\right)\frac{1}{1+e^{-\left(\phi_{y}+\U\left(\xi-y\right)+\lambda\right)}}.\label{eq:Saddle_point_phi_final_g=0}\\
   f&=\int dx\, \frac{1}{1+e^{-\left(\phi_{x}+\U\left(\xi-r_{i}\right)+\lambda\right)}},
\end{align}
the latter two corresponding to eqs. (12) - (13) in \cite{Rosay13_062813}. In particular, all auxiliary fields have a well-behaved limit for $g\rightarrow 0$.
Furthermore, from eq. (\ref{eq:Implicit_def_auxiliary_var}), we obtain the derivatives of the auxiliary variables with repect to $g$ to be given by
\begin{align}
	\frac{\partial}{\partial g} 
	\begin{pmatrix}
    		\left\{\phi_{x}\right\}_{x}\\
    		\lambda\\
    		q
    	\end{pmatrix}
	= \left(\frac{\partial^2}{\partial \left[\left\{\phi_{x}\right\}_{x},\lambda,q\right]^{2}} G\left[q,\boldsymbol{\phi},\lambda\right]\right)^{-1}
	\frac{\partial^{2}}{\partial g \partial \left[\left\{\phi_{x}\right\}_{x},\lambda,q\right]} G_{g} \left[q,\boldsymbol{\phi},\lambda\right].
\end{align}
Further differentiating $\frac{\partial}{\partial \left[\left\{\phi_{x}\right\}_{x},\lambda,q\right]}G_{g}\left[q,\boldsymbol{\phi},\lambda\right]$ with respect to $g$ yields
\begin{align}
	\frac{\partial^2}{\partial g \partial\left[\left\{\phi_{x}\right\}_{x},\lambda,q\right]} G_{g} \left[q,\boldsymbol{\phi},\lambda\right] =
    	\begin{pmatrix}
			\left\{\int dy\, K_{xy}\int Dt_{y}\, t_{y}  \sqrt{2q} \cdot m_{y}\left(1 - 2m_{y}\right)\right\}_{x}\\
    		\int dy\, \int{\cal D}t_{y}\, t_{y} \sqrt{2q} \cdot m_{y}\left(1 - 2m_{y}\right)\\
    		\int dx\,\int Dt_{y}\, t_{y}  \sqrt{2q} \cdot m_{y}\left(1 - 2m_{y}\right) \left(1 - m_{y}\right)
    	\end{pmatrix} \overset{g=0}{=} 0,\label{eq:Derivative_SP_equations_wrt_g}  
\end{align}
with $m_{x}$ as introduced in eq. (\ref{eq:Def_mx}). The last equality in eq.  (\ref{eq:Derivative_SP_equations_wrt_g}) holds because for $g=0$, $m_{x}$ is independent of $t_{x}$ and the remaining $t_{x}$-integrand is antisymmetric. To obtain the derivatives of the order parameters at $g=0$, we therefore only have to check that differentiating $\partial_{\left\{\phi_{x}\right\}_{x},\lambda,q}G$ with respect to $q, \phi_{x}$ and $\lambda$ once more yields a regular Hessian. We obtain
\begin{align}
	 \frac{\partial^2}{\partial \left[\left\{\phi_{x}\right\}_{x},\lambda,q\right]^{2}} G_{g} \left[q,\boldsymbol{\phi},\lambda\right]
	=\begin{pmatrix}
    		-\left(K^{-1}\right)_{xy} +  \delta_{xy}  \kv_{y} & \kv_{x} & g^{2} \ks_{x} \\
    		 \kv_{y} & \int dz\, \kv_{z} & g^{2} \int dz\, \ks_{z}\\
    		 g^{2} \ks_{y}  & g^{2} \int dz\, \ks_{z} & g^{2} + g^{4}\int dz\, \kk_{z}
    	\end{pmatrix},
\end{align}
where we have omitted the $t_{x}$-dependence of $m_{x}$ for brevity and have introduced the higher-order cumulants
\begin{align}
    \ks_{x} \coloneqq &\frac{\partial^{2} m_{x}}{\partial \phi_{x}^{2}} = \int{\cal D}t\, m_{x}\left(1 - 2m_{x}\right) \left(1 - m_{x}\right) \\
    \kk_{x} \coloneqq &\frac{\partial^{3} m_{x}}{\partial \phi_{i}^{3}} = \int{\cal D}t\, m_{x}\left(1 - m_{x}\right)\left(1 - 6m_{x} + 6m_{x}^{2}\right),
\end{align}
(the fourth-order one for later use).
It is not apparent why the Hessian should have a zero mode - indeed, this would mean in particular that the saddle-point approximation is not well-defined. So as long as we trust the saddle-point approximation, we also know that the derivatives of the order-parameters with respect to $g$ vanish for $g=0$. Also, in appendix  \ref{sec:CompFisherInfo}, we numerically confirm that the Hessian does not have zero modes for $g\rightarrow 0$.

\section{\label{sec:CompFisherInfo}Computing the Fisher information}
\begin{figure}
	\centering
    \subfloat{{\includegraphics[width=.5\textwidth]{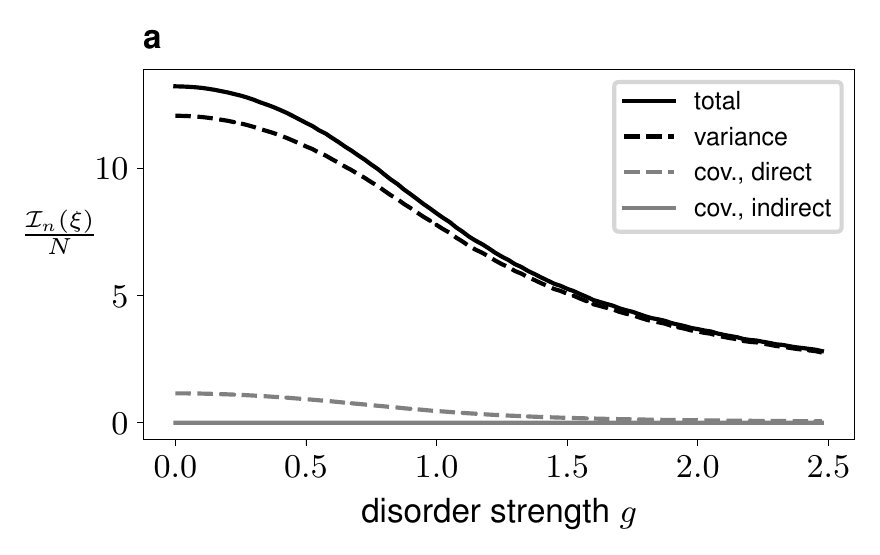} }}
    \subfloat{{\includegraphics[width=.5\textwidth]{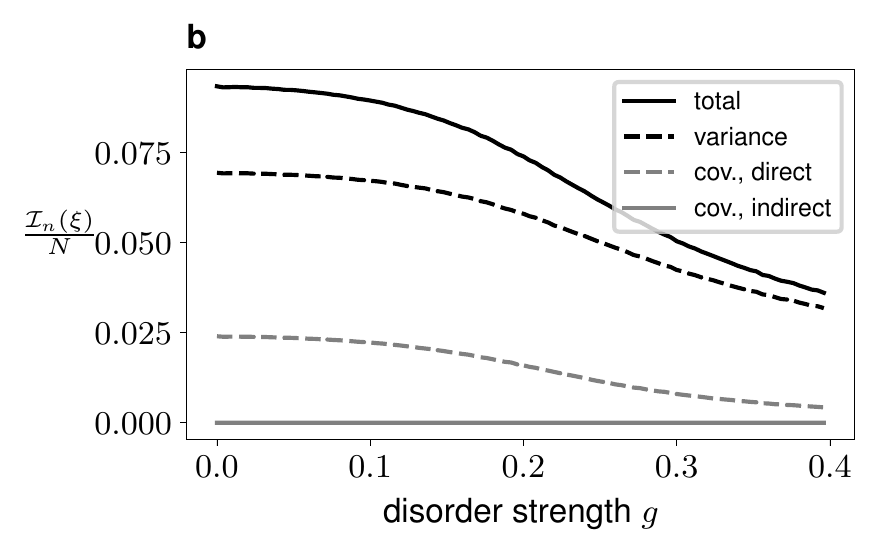} }}
     \caption{The Fisher information per neuron in dependence of the disorder,  contributions from different parts of covariance as defined in eq. (\ref{eq:Covariance_contributions}). Parameters as in fig.  (\ref{fig:FisherInfo_vs_g_MC_comparison}).}
	\label{fig:FisherInfo_vs_g_theo_different_contributions}
\end{figure}

First, we convince ourselves that computing the Fisher information simply amounts to computing the covariance matrix. Consider the probability distribution for the
neural network state $\boldsymbol{n}$, conditioned on the stimulus
$\xi$, $P\left(\boldsymbol{n}|\xi\right)$, given by
\[
P\left(\boldsymbol{n}|\xi\right)=\frac{1}{{\cal Z}_{J}\left(\xi\right)}e^{\sum_{i}n_{i}\U\left(\xi-r_{i}\right)+\sum_{i<j}\left(J_{ij} + K_{ij}\right) n_{i}n_{j}},
\]
and
\[
{\cal Z}_{J}\left(\xi\right)=\sum_{\boldsymbol{n}}e^{\sum_{i}n_{i}\U\left(\xi-r_{i}\right)+\sum_{i<j}\left(J_{ij} + K_{ij}\right) n_{i}n_{j}}.
\]
For the Fisher information, we need the second derivative of the
logarithm of $P$ with respect to $\xi$:
\begin{align*}
-\frac{\partial^{2}}{\partial\xi^{2}}\ln\left(P\left(\boldsymbol{n}|\xi\right)\right)= & -\frac{\partial^{2}}{\partial\xi^{2}}\sum_{i}n_{i}\U\left(\xi-r_{i}\right)+\frac{\partial^{2}}{\partial\xi^{2}}\ln{\cal Z}_{J}\left(\xi\right)\\
= & -\frac{\partial}{\partial\xi}\sum_{i}n_{i}\U^{\prime}\left(\xi-r_{i}\right)+\frac{\partial}{\partial\xi}\frac{\frac{\partial}{\partial\xi}{\cal Z}_{J}\left(\xi\right)}{{\cal Z}_{J}\left(\xi\right)}\\
= & -\sum_{i} n_{i}\U^{\prime\prime}\left(\xi-r_{i}\right)+\frac{\frac{\partial^{2}}{\partial\xi^{2}}{\cal Z}_{J}\left(\xi\right)}{{\cal Z}_{J}\left(\xi\right)}-\left(\frac{\frac{\partial}{\partial\xi}{\cal Z}_{J}\left(\xi\right)}{{\cal Z}_{J}\left(\xi\right)}\right)^{2}.
\end{align*}
Upon averaging over the neurons states of the neuron $n_{i}$, we obtain
\begin{align}
    {\cal I}_{\boldsymbol{n}}\left(\xi\right)= & \left\langle -\frac{1}{{\cal Z}_{J}\left(\xi\right)}\sum_{\boldsymbol{n}}e^{\sum_{i}n_{i}\U\left(\xi-r_{i}\right)+\sum_{i<j}J_{ij}n_{i}n_{j}}\sum_{i}n_{i}\U^{\prime\prime}\left(\xi-r_{i}\right)\right.\nonumber \\
 & +\frac{1}{{\cal Z}_{J}\left(\xi\right)}\sum_{\boldsymbol{n}}e^{\sum_{i}n_{i}\U\left(\xi-r_{i}\right)+\sum_{i<j}J_{ij}n_{i}n_{j}}\left[\sum_{i}n_{i}\U^{\prime\prime}\left(\xi-r_{i}\right)+\left(\sum_{i}n_{i}\U^{\prime}\left(\xi-r_{i}\right)\right)^{2}\right]\nonumber \\
 & \left.-\left(\frac{1}{{\cal Z}_{J}\left(\xi\right)}\sum_{\boldsymbol{n}}e^{\sum_{i}n_{i}\U\left(\xi-r_{i}\right)+\sum_{i<j}J_{ij}n_{i}n_{j}}\sum_{i}n_{i}\U^{\prime}\left(\xi-r_{i}\right)\right)^{2}\right\rangle_{J}\label{eq:Fisher_information_total}\\
 = & \sum_{i,j}\U^{\prime}\left(\xi-r_{i}\right)\left\langle \left[\left\langle n_{i}n_{j}\right\rangle _{\boldsymbol{n}}-\left\langle n_{i}\right\rangle _{\boldsymbol{n}}\left\langle n_{j}\right\rangle _{\boldsymbol{n}}\right]\right\rangle _{J}\U^{\prime}\left(\xi-r_{j}\right),
\end{align}
where we have used the usual thermal average
\begin{align}
    \left\langle f\left(\boldsymbol{n}\right)\right\rangle _{\boldsymbol{n}}:=\frac{1}{{\cal Z}_{J}\left(\xi\right)}\sum_{\boldsymbol{n}}f\left(\boldsymbol{n}\right)e^{\sum_{i}n_{i}\U\left(\xi-r_{i}\right)+\sum_{i<j}J_{ij}n_{i}n_{j}}.
\end{align}
As indicated before, to determine the Fisher information, we therefore just have to compute the covariance matrix, which we achieve by differentiating the cumulant-generating functional twice with respect to $\boldsymbol{h}$, considering all indirect dependencies via the auxiliary fields (evaluated at their respective saddle-point values). Taking into account both the fixed total activity and the disorder, the cumulant-generating functional is given by eq. (\ref{eq:Def_G_cum_gen_fct}). Formally differentiating this expression yields
\begin{align}
    \frac{d^{2}}{d \boldsymbol{h}^{2}}\left\langle W_{f}\left(\boldsymbol{h}\right)\right\rangle_{J}= & \frac{\partial^{2}G}{\partial\boldsymbol{h}^{2}}+2\frac{\partial^{2}G}{\partial\boldsymbol{h}\partial\boldsymbol{\phi}}\frac{\partial\boldsymbol{\phi}}{\partial\boldsymbol{h}}+2\frac{\partial^{2}G}{\partial\boldsymbol{h}\partial q}\frac{\partial q}{\partial\boldsymbol{h}} + 2\frac{\partial^{2}G}{\partial\boldsymbol{h}\partial \lambda}\frac{\partial \lambda}{\partial\boldsymbol{h}} \\
 & +\frac{\partial^{2}G}{\partial q^{2}}\left(\frac{\partial q}{\partial\boldsymbol{h}}\right)^{2}+\frac{\partial^{2}G}{\partial \lambda^{2}}\left(\frac{\partial \lambda}{\partial\boldsymbol{h}}\right)^{2} +\frac{\partial\boldsymbol{\phi}}{\partial\boldsymbol{h}}\frac{\partial^{2}G}{\partial\boldsymbol{\phi}^{2}}\frac{\partial\boldsymbol{\phi}}{\partial\boldsymbol{h}}\\
 &+2\frac{\partial^{2}G}{\partial q\partial\lambda}\frac{\partial q}{\partial\boldsymbol{h}}\frac{\partial\lambda}{\partial\boldsymbol{h}}
 +2\frac{\partial^{2}G}{\partial q\partial\boldsymbol{\phi}}\frac{\partial q}{\partial\boldsymbol{h}}\frac{\partial\boldsymbol{\phi}}{\partial\boldsymbol{h}}+2\frac{\partial^{2}G}{\partial \lambda\partial\boldsymbol{\phi}}\frac{\partial \lambda}{\partial\boldsymbol{h}}\frac{\partial\boldsymbol{\phi}}{\partial\boldsymbol{h}}.\label{eq:Diff_cfg_indirect_contr}
\end{align}
We obtain the derivatives of $q$ and $\boldsymbol{\phi}$ by taking the total derivatives of their defining equations, i.e. \eqref{eq:Saddle_point_q} and  \eqref{eq:Saddle_point_phi_final}, which yields
\begin{align}
    0=\frac{d}{d\boldsymbol{h}}\frac{\partial}{\partial q}G\left(\boldsymbol{h},\boldsymbol{\phi},q,\lambda\right) & =\frac{\partial^{2}G}{\partial q^{2}}\frac{\partial q}{\partial\boldsymbol{h}}
    +\frac{\partial^{2}G}{\partial\lambda\partial q}\frac{\partial\lambda}{\partial\boldsymbol{h}}
    +\frac{\partial^{2}G}{\partial\boldsymbol{\phi}\partial q}\frac{\partial\boldsymbol{\phi}}{\partial\boldsymbol{h}}
    +\frac{\partial^{2}G}{\partial q\partial\boldsymbol{h}}\\
    0=\frac{d}{d\boldsymbol{h}}\frac{\partial}{\partial \lambda}G\left(\boldsymbol{h},\boldsymbol{\phi},q,\lambda\right) & =\frac{\partial^{2}G}{\partial q\partial \lambda}\frac{\partial q}{\partial\boldsymbol{h}}
    + \frac{\partial^{2}G}{\partial \lambda^{2}}\frac{\partial \lambda}{\partial\boldsymbol{h}}
    +\frac{\partial^{2}G}{\partial\boldsymbol{\phi}\partial \lambda}\frac{\partial\boldsymbol{\phi}}{\partial\boldsymbol{h}}
    +\frac{\partial^{2}G}{\partial \lambda\partial\boldsymbol{h}}\\
0=\frac{d}{d\boldsymbol{h}}\frac{\partial}{\partial\boldsymbol{\phi}}G\left(\boldsymbol{h},\boldsymbol{\phi},q,\lambda\right) & =
\frac{\partial^{2}G}{\partial q\partial\boldsymbol{\phi}}\frac{\partial q}{\partial\boldsymbol{h}}
+\frac{\partial^{2}G}{\partial \lambda\partial\boldsymbol{\phi}}\frac{\partial \lambda}{\partial\boldsymbol{h}}
+\frac{\partial^{2}G}{\partial\boldsymbol{\phi}^{2}}\frac{\partial\boldsymbol{\phi}}{\partial\boldsymbol{h}}+\frac{\partial^{2}G}{\partial\boldsymbol{\phi}\partial\boldsymbol{h}},
\end{align}
so that we obtain after inserting into \eqref{eq:Diff_cfg_indirect_contr}
\begin{align}
    \frac{d^{2}}{d \boldsymbol{h}^{2}}\left\langle W_{f}\left(\boldsymbol{h}\right)\right\rangle_{J} = \frac{\partial^{2}G}{\partial\boldsymbol{h}^{2}}-\begin{pmatrix}\frac{\partial^{2}G}{\partial\boldsymbol{h}\partial\boldsymbol{\phi}}\\
\frac{\partial^{2}G}{\partial\boldsymbol{h}\partial q}\\
\frac{\partial^{2}G}{\partial\boldsymbol{h}\partial\lambda}
\end{pmatrix}^{\mathrm{T}}\begin{pmatrix}\frac{\partial^{2}G}{\partial\boldsymbol{\phi}^{2}} & \frac{\partial^{2}G}{\partial\boldsymbol{\phi}\partial q} & \frac{\partial^{2}G}{\partial\boldsymbol{\phi}\partial\lambda}\\
\frac{\partial^{2}G}{\partial q\partial\boldsymbol{\phi}} & \frac{\partial^{2}G}{\partial q^{2}} & \frac{\partial^{2}G}{\partial q\partial\lambda}\\
\frac{\partial^{2}G}{\partial\lambda\partial\boldsymbol{\phi}} & \frac{\partial^{2}G}{\partial\lambda\partial q} & \frac{\partial^{2}G}{\partial\lambda^{2}}
\end{pmatrix}^{-1}\begin{pmatrix}\frac{\partial^{2}G}{\partial\boldsymbol{\phi}\partial\boldsymbol{h}}\\
\frac{\partial^{2}G}{\partial q\partial\boldsymbol{h}}\\
\frac{\partial^{2}G}{\partial\lambda\partial\boldsymbol{h}}
\end{pmatrix}.
\end{align}
In order to compactly write down the entries of the matrix and the vectors above, we introduce the effective local connectivity
\begin{align}
    \left(\Jspace_{\mathrm{eff}}^{-1}\right)_{xy}  \coloneqq -\frac{\partial^{2}G}{\partial\phi_{x}\partial\phi_{y}} \Leftrightarrow\left[\left(\frac{\partial^{2}G}{\partial\phi\partial\phi}\right)^{-1}\right]_{xy} = -\left(\Jspace_{\mathrm{eff}}\right)_{xy},
\end{align}
which fulfills the Dyson equation whose concrete form we obtain by performing the derivatives of $G$ explicitly:
\begin{align}
    \left(\Jspace_{\mathrm{eff}}^{-1}\right)_{xy} & =\left(\Jspace^{-1}\right)_{xy}-\delta_{xy} \kv_{x}\\
    \Leftrightarrow \Jspace_{xy}^{\mathrm{eff}} & =\Jspace_{xy} + \int\,  \Jspace_{xz} \kv_{z} \Jspace_{zy}^{\mathrm{eff}}.
\end{align}
Using the identities derived in appendix \ref{sec:MatVecCalc}, in particular eq. (\ref{eq:Sandwiched_inverse_block_matrix}), we can note the final form of the covariance matrix:
\begin{align}
     C =  \Kv + \Kv \Jspace_{\mathrm{eff}} \Kv - \left(\mathbb{1}_{N}+\Kv \Jspace_{\mathrm{eff}}\right)
     \begin{pmatrix}g\ksvec, & \kvvec
     \end{pmatrix} 
    S^{-1}
    \begin{pmatrix}
    	g\ksvec\\
    	\kvvec
    \end{pmatrix}\left(\mathbb{1}_{N}+\Jspace_{\mathrm{eff}}\Kv\right),\label{eq:Covariance_quasifinal}
\end{align}
where $\Kv$ is the diagonal matrix with the disorder-averaged variances $\kv_{i}$ and
\begin{align}
	S =\begin{pmatrix}
    		N+g^{2}\sum_{i}\kk_{i} & g\sum_{i}\ks_{i}\\
    		g\sum_{i}\ks_{i} & \sum_{i}\kv_{i}
    	\end{pmatrix}+\begin{pmatrix}g\ksvec\\
    	\kvvec
    	\end{pmatrix}\Jspace_{\mathrm{eff}}
    	\begin{pmatrix}
    		g\ksvec, & \kvvec
    	\end{pmatrix}
\end{align}
The Fisher information, finally is then given by
\begin{align}
	{\cal I}_{\boldsymbol{n}}\left(\xi\right)=\sum_{x,y}\U^{\prime}\left(\xi-x\right)  C_{xy}\U^{\prime}\left(\xi-y\right).
\end{align}
The last term from \ref{eq:Covariance_quasifinal} contributes to the Fisher information with a term containing (twice) the expression
\begin{eqnarray}
	    \begin{pmatrix}
    	g\ksvec\\
    	\kvvec
    \end{pmatrix}\left(\mathbb{1}_{N}+\Jspace_{\mathrm{eff}}\Kv\right) \boldsymbol{U}^{\prime}
\end{eqnarray}
The space-dependence of the contribution from the covariance is mostly determined by the shape of $U$ (multiplications by $\Jspace$ or $\Jspace_{\mathrm{eff}}$ merely smear it out), so that multiplication with its derivative with $U^{\prime}$ is well approximated by a spatial derivative and summation over space and therefore yields a contribution close to $0$. This part of the covariance therefore only yields subleading contributions to the Fisher information (see fig. (\ref{fig:FisherInfo_vs_g_theo_different_contributions})) and we can neglect it in the analysis. This makes sense because it emerges from the source-dependence of $q$ and $\lambda$, the auxiliary variables representing the disorder and the global inhibition, which are global quantities. It is therefore expected that their contribution to the spatial information is negligible.

\subsection{Analysis of the covariance matrix}
Having an analytical expression for the covariance matrix at hand, we can investigate its behavior for special cases, in particular around $g=0$. Because it depends on $g$ only via the cumulants $\kv$, $\ks$ and $\kk$, we primarily have to examine their behavior near $g=0$. We observe that 
\begin{align}
	\frac{d}{d g} \kv_{x} &= \frac{\partial}{\partial g} \kv_{x} + \frac{\partial \kv_{x}}{\partial q} \frac{\partial q}{\partial g} + \int dy\, \frac{\partial \kv_{x}}{\partial \phi_{y}} \frac{\partial \phi_{y}}{\partial g} + \frac{\partial \kv_{x}}{\partial \lambda} \frac{\partial \lambda}{\partial g} \\&= \frac{\partial}{\partial g} \int{\cal D}t\, m_{x} \left(1- m_{x}\right) = \int{\cal D}t\, t \sqrt{2q}\, m_{x}\left(1-3m_{x}+2m_{x}^{2}\right)   \overset{g=0}{=} 0, 			   
\end{align}
where we have used the result from appendix (\ref{sec:Analysis_SP_eq}) that the derivatives of the auxiliary variables with respect to $g$ vanish as $g$ goes to $0$. Again because $m_{x}$ does not depend on $t_{x}$ for $g=0$ and the remaining integral over $t_{x}$ is antisymmetric, it also yields $0$. With the same argument, the derivatives of the other cumulants vanish as well. Therefore, the linear orders of all $g$-dependent quantities, that the covariance $C$ depends on, vanish. Thus, the derivatives of the covariances and of the Fisher information equal $0$ for $g=0$ as well, as apparent from the plots in figure (\ref{fig:FisherInfo_vs_g_MC_comparison}).

\subsection{Relating inputs and tuning curves by means of $\Jspace_{\mathrm{eff}}$ \label{sec:Derivation_input_tuning_curve_Keff}}
Without disorder, the tuning curve $\Tc$ in the thermodynamic limit is given by
\begin{align}
    \Tc_{x} &= \int dy\, K\left(x-y\right)\frac{1}{1+e^{-\left(\phi_{y}+\U\left(\xi-y\right)+\lambda\right)}}\\
    \phi_{y} &= \int dx\, K\left(y - z\right) \Tc_{z}
\end{align}
and therefore, we can write for its derivative
\begin{align}
    \boldsymbol{\Tc}^{\prime}	&=\boldsymbol{\Tc}\left(1 - \boldsymbol{\Tc}\right)\left(\boldsymbol{\U}^{\prime}+K \boldsymbol{\Tc}^{\prime}\right)\\
    \Leftrightarrow\left[1 - \boldsymbol{\Tc}\left(1 - \boldsymbol{\Tc}\right)K\right]\boldsymbol{\Tc}^{\prime}	&= \boldsymbol{\Tc}\left(1-\boldsymbol{\Tc}\right)\boldsymbol{\U}^{\prime}\\
    \boldsymbol{\Tc}^{\prime}	&= \left(1-VK\right)^{-1}V \boldsymbol{\U}^{\prime}=\left(V^{-1}-K\right)^{-1}\boldsymbol{\U}^{\prime}\\
    \Leftrightarrow \boldsymbol{\Tc}^{\prime}	&=V\left(1+K\left(V^{-1}-K\right)^{-1}\right)\boldsymbol{\U}^{\prime},
    \label{eq:Rearrange_relation_tuning_curve_input}
\end{align}    
where we have abbreviated $V_{ij} = \delta_{ij}f_{i}\left(1 - f_{i}\right)$. We furthermore have
\begin{align}
    \Jspace_{\mathrm{eff}} & =\Jspace + \Jspace V \Jspace_{\mathrm{eff}}\\
\Leftrightarrow\left(1 - \Jspace V\right)\Jspace_{\mathrm{eff}} & = \Jspace\\
\Leftrightarrow \Jspace_{\mathrm{eff}} & =\left(1 - \Jspace V\right)^{-1}\Jspace = V^{-1}\left(V^{-1} - \Jspace\right)^{-1}\Jspace\\
\Leftrightarrow \Jspace_{\mathrm{eff}} & = \Jspace\left(V^{-1}-\Jspace\right)^{-1}V^{-1}\\
\Leftrightarrow \Jspace_{\mathrm{eff}}V & = \Jspace\left(V^{-1} - \Jspace\right)^{-1},
\end{align}
where we obtained the second-to-last equivalence by transposing. Inserting this expression into eq. (\ref{eq:Rearrange_relation_tuning_curve_input}), we arrive at eq. (\ref{eq:Tuning_curve_with_Keff}).
\section{Matrix-vector calculus}\label{sec:MatVecCalc}
\subsection{Inversion of a matrix with blocks on the diagonal of the sizes $N$
and $M$}
Assume we have a matrix of the form
\begin{align}
    U:=\begin{pmatrix}A & b\\
    b^{\mathrm{T}} & a
    \end{pmatrix},
\end{align}
where 
\begin{equation}
    A\in\mathbb{R}^{N\times N},\ b\in\mathbb{R}^{N\times M},\ a\in\mathbb{R}^{M\times M},
\end{equation}
$a$ and $A$ are symmetric and $A$ is invertible. To invert it, we make the ansatz
\begin{equation}
    V:=\begin{pmatrix}C & d\\
    d^{\mathrm{T}} & c
    \end{pmatrix}.\label{eq:Inverse_block_plus_one_matrix}
\end{equation}
Multiplying $U$ and $V$, we obtain the conditions
\begin{align}
    AC+bd^{\mathrm{T}} & =\mathbb{1}_{N}\label{eq:First_one_matrix_eq}\\
    Ad+bc & =0\label{eq:First_zero_matrix_eq}\\
    b^{\mathrm{T}}C+ad^{\mathrm{T}} & =0\label{eq:Second_zero_matrix_eq}\\
    b^{\mathrm{T}}d+ac & =\mathbb{1}_{M}\label{eq:Second_one_matrix_eq}
\end{align}
Solving \eqref{eq:First_zero_matrix_eq} for $d$ and inserting into
\eqref{eq:Second_one_matrix_eq}, we obtain
\begin{align*}
    c & =\left(a-b^{\mathrm{T}}A^{-1}b\right)^{-1}\\
    \mathrm{and}\ d & =-A^{-1}b\left(a-b^{\mathrm{T}}A^{-1}b\right)^{-1}.
\end{align*}
Solving \eqref{eq:First_one_matrix_eq} for $C$ and inserting the results
gained until here, we obtain
\[
C=A^{-1}+A^{-1}b\left(a-b^{\mathrm{T}}A^{-1}b\right)^{-1}\left(A^{-1}b\right)^{\mathrm{T}}.
\]
Plugging these results into the left-hand side of \eqref{eq:Second_zero_matrix_eq},
which we did not use so far, we obtain
\begin{align*}
 & b^{\mathrm{T}}\left(A^{-1}+A^{-1}b\left(a-b^{\mathrm{T}}A^{-1}b\right)^{-1}\left(A^{-1}b\right)^{\mathrm{T}}\right)-a\left[A^{-1}b\left(a-b^{\mathrm{T}}A^{-1}b\right)^{-1}\right]^{\mathrm{T}}\\
= & \left(\left(A^{-1}b\right)^{\mathrm{T}}+\left(b^{\mathrm{T}}A^{-1}b-a+a\right)\left(a-b^{\mathrm{T}}A^{-1}b\right)^{-1}\left(A^{-1}b\right)^{\mathrm{T}}\right)-a\left(a-b^{\mathrm{T}}A^{-1}b\right)^{-1}\left(A^{-1}b\right)^{\mathrm{T}}\\
= & a\left(a-b^{\mathrm{T}}A^{-1}b\right)^{-1}\left(A^{-1}b\right)^{\mathrm{T}}-a\left(a-b^{\mathrm{T}}A^{-1}b\right)^{-1}\left(A^{-1}b\right)^{\mathrm{T}}=0,
\end{align*}
therefore our ansatz is consistent. Summarizing, we can write the
inverse of $U$ as 
\begin{equation}
U^{-1}=\begin{pmatrix}A^{-1} & 0\\
0 & 0
\end{pmatrix}+\begin{pmatrix}-A^{-1}b\\
\mathbb{1}_{M}
\end{pmatrix}\left(a-b^{\mathrm{T}}A^{-1}b\right)^{-1}\begin{pmatrix}-\left(A^{-1}b\right)^{\mathrm{T}} & ,\mathbb{1}_{M}\end{pmatrix}.\label{eq:Final_expression_inverse_block_matrix}
\end{equation}

\subsection{Vector-matrix-vectors multiplication}

Calculating cross-covariances, we are interested in calculating objects
of the type
\[
\begin{pmatrix}B & b\end{pmatrix}\begin{pmatrix}A & b\\
b^{\mathrm{T}} & a
\end{pmatrix}^{-1}\begin{pmatrix}B\\
b^{\mathrm{T}}
\end{pmatrix}.
\]
Making use of \eqref{eq:Final_expression_inverse_block_matrix},
we then obtain
\begin{align}
 & \begin{pmatrix}B & b\end{pmatrix}\begin{pmatrix}A & b\\
b^{\mathrm{T}} & a
\end{pmatrix}^{-1}\begin{pmatrix}B\\
b^{\mathrm{T}}
\end{pmatrix}\nonumber \\
= & BA^{-1}B+\left(\mathbb{1}_{N}-BA^{-1}\right)b\left(a-b^{\mathrm{T}}A^{-1}b\right)^{-1}b^{\mathrm{T}}\left(\mathbb{1}_{N}-A^{-1}B\right)\label{eq:Sandwiched_inverse_block_matrix}
\end{align}
\section{Relating Fisher and mutual information for uncoupled neurons with inhomogeneously distributed place fields\label{sec:MutualInfo_FisherInfo_independent}}
Here, we consider independent neurons, but allow variability in the tuning curves $\Tc$. The probability distribution of the neural population is then given by
\begin{equation}
	P\left(\boldsymbol{n}\right) = \prod_{i=1}^{N} \left[n_{i} \Tc_{i}\left(\xi\right) + \left(1 - n_{i}\right) \left(1 - \Tc_{i}\left(\xi\right)\right)\right] P_{\xi}\left(\xi\right). 
\end{equation}
To compute the mutual information, we first need to compute the entropy of this distribution, which is given by
\begin{align}
  h_{\mathrm{uncond}}\nonumber 
=  -\lim_{N\rightarrow\infty}\frac{1}{N}\sum_{\boldsymbol{n}}\left\langle P\left(\boldsymbol{n}\right)\ln\left(P\left(\boldsymbol{n}\right)\right)\right\rangle_{\Tc},
\end{align}
where we denote by $\left\langle \dots \right\rangle_{\Tc}$ the average over the variability of the tuning curves. The tricky part here is that we average over a logarithm, a complication that we deal with by introducing replicas ($n+1$ in this case because of the prefactor $P\left(\boldsymbol{n}\right)$, compare \cite{Nadal93_295}), which leads to
\begin{align}
h_{\mathrm{uncond}}
= & -\lim_{N\rightarrow\infty}\frac{1}{N}\lim_{k\rightarrow 0}\frac{1}{k}\left\{ \int\prod_{\alpha=0}^{k}\left(d\xi_{\alpha}P_{\xi}\left(\xi_{\alpha}\right)\right)\,\left(\left\langle \left[\prod_{\alpha=0}^{k}\Tc\left(\xi_{\alpha}\right)+\prod_{\alpha=0}^{k}\left(1 - \Tc\left(\xi_{\alpha}\right)\right)\right]\right\rangle_{\Tc}\right)^{N}-1\right\} \\
= & -\lim_{N\rightarrow\infty}\frac{1}{N}\lim_{k\rightarrow 0}\frac{1}{k}\left\{ \int\prod_{\alpha=0}^{k}\left(d\xi_{\alpha}P_{\xi}\left(\xi_{\alpha}\right)\right)\,\left(G_{\Tc}\left(\boldsymbol{\xi}\right)\right)^{N}-1\right\},\label{eq:calc_h_uncond_with_G}
\end{align}
where we have introduced
\[
G_{\Tc}\left(\boldsymbol{\xi}\right)\coloneqq\left\langle \left[\prod_{\alpha=0}^{n}\Tc\left(\xi_{\alpha}\right)+\prod_{\alpha=0}^{n}\left(1 - \Tc\left(\xi_{\alpha}\right)\right)\right]\right\rangle_{\Tc}.
\]
An obvious idea is now to evaluate eq. (\ref{eq:calc_h_uncond_with_G}) in saddle-point approximation, as also shown in \cite{Brunel98_1731, Sompolinsky01_051904}. This is indeed what we will do, but with a small twist because one of the eigenvalues of the Hessian of $G_{\Tc}$ vanishes for $n\rightarrow 0$. However, this replicon mode can be identified to be the one corresponding to the replica-symmetric direction. This allows us to transform the $n+1$-dimensional integral over the $\xi_{\alpha}$ such that the first coordinate corresponds to the replica-symmetric direction $\left(1,\dots, 1\right)$ and the other $n$ are orthogonal to it. Like this, we can  perform the integral over the first coordinate exactly and only the orthogonal directions are evaluated in saddle-point approximation. Having determined the unconditioned entropy in this way, we obtain the mutual information $\mathrm{MI} = h_{\mathrm{uncond}} - h_{\mathrm{cond}}$ by subtracting
\begin{equation}
	h_{\mathrm{cond}} = - \lim_{N\rightarrow\infty}\frac{1}{N}\sum_{\boldsymbol{n}}\int d\xi\,\left\langle P\left(\boldsymbol{n}|\xi\right)\ln\left(P\left(\boldsymbol{n}|\xi\right)\right)\right\rangle_{\Tc}
\end{equation}
from the unconditioned entropy. Performing the limit of $k\rightarrow 0$ in eq. (\ref{eq:calc_h_uncond_with_G}), we see that, to zeroth order, $h_{\mathrm{uncond}}$ equals $h_{\mathrm{cond}}$, so that the mutual information is, to first order, given by the one-loop correction
\begin{equation}
	I_{\mathrm{MI}} = \int d\xi P_{\xi}\left(\xi\right)\, \left[\frac{1}{2} \ln\left(-\frac{N\lambda^{1,k=0}_{\Tc}\left(\xi\right)}{2\pi}\right) - \frac{1}{2} - \ln\left(P_{\xi}\left(\xi\right)\right)\right] + \mathcal{O}\left(\frac{1}{N}\right),
	\label{eq:RelMI_FI_with_EV}
\end{equation}
where $\lambda^{1,k=0}_{\Tc}$ is the $n$-fold degenerate eigenvalue of $G_{\Tc}$ at $k=0$ (see below).
Note that our computation neither requires the introduction of helping fields to perform the average over the state space of the neural population, as \cite{Brunel98_1731}, nor do we assume it to be normally distributed, as in \cite{Sompolinsky01_051904}. However, in return, we are assuming the neurons to be independent, which limits the applicability of our approach.\\
What is left to do is the computation of the Hessian of $G_{\Tc}$. On the replica-symmetric line, we only have two values for its entries, the diagonal and the off-diagonal. We calculate
\begin{align}
\left.\frac{\partial^{2}G_{\Tc}}{\partial\xi_{\alpha}^{2}}\right|_{\xi_{0}=\dots=\xi_{n}=\xi}
 =&\sum_{n=0,1}\left.\left\langle \prod_{\gamma=0,\gamma\neq\alpha}^{k}\left[n \Tc\left(\xi_{\gamma}\right) \left(1 - n\right) \left( 1- \Tc\left(\xi_{\gamma}\right)\right)\right]\left(2n-1\right) \Tc^{\prime\prime}\left(\xi_{\alpha}\right)\right\rangle_{r}\right|_{\xi_{0}=\dots=\xi_{n}=\xi}\\
 =&\sum_{n=0,1}\left(2n-1\right) \left\langle \left[n\Tc\left(\xi\right)+ \left(1 - n\right) \left(1 - \Tc\left(\xi\right)\right)\right]^{k}\Tc^{\prime\prime}\left(\xi\right)\right\rangle_{r}\\
 \overset{k=0}{=} &0. 
 \label{eq:HessGTc_diagonal}
\end{align}
and
\begin{align}
 & \left.\frac{\partial^{2}G_{\Tc}}{\partial\xi_{\alpha}\partial\xi_{\beta}}\right|_{\xi_{0}=\dots=\xi_{n}=\xi}\\
= & \sum_{n=0,1}\left.\left\langle \prod_{\gamma=0,\gamma\neq\alpha,\beta}^{k}\left[n\Tc\left(\xi_{\gamma}\right) + \left(1 - n\right) \left(1 - \Tc\left(\xi_{\gamma}\right)\right)\right]\left(2n-1\right)^{2}\Tc^{\prime}\left(\xi_{\alpha}\right)\Tc^{\prime}\left(\xi_{\beta}\right)\right\rangle_{r}\right|_{\xi_{0}=\dots=\xi_{n}=\xi}\\
= & \sum_{n=0,1}\left\langle \left[n\Tc\left(\xi\right)+\left(1-n\right)\left(1-\Tc\left(\xi\right)\right) \right]^{k-1}\left[\Tc^{\prime}\left(\xi\right)\right]^{2}\right\rangle_{r}\\
\overset{k=0}{=} & \left\langle \left[\frac{1}{\Tc\left(\xi\right)} + \frac{1}{1-\Tc\left(\xi\right)} \right]\left[\Tc^{\prime}\left(\xi\right)\right]^{2}\right\rangle_{r} = \left\langle \frac{\left[\Tc^{\prime}\left(\xi\right)\right]^{2}}{\Tc\left(\xi\right)\left(1-\Tc\left(\xi\right)\right)}\right\rangle_{r}.
\label{eq:HessGTc_offdiagonal}
\end{align}
The eigenvalues of the Hessian of $G_{\Tc}$ are given by
\begin{align*}
\lambda_{0}\left(\xi\right) & =\left.\frac{\partial^{2}G}{\partial\xi_{\alpha}^{2}}\right|_{\xi_{0}=\dots=\xi_{n}=\xi}+n\left.\frac{\partial^{2}G}{\partial\xi_{\alpha}\partial\xi_{\beta}}\right|_{\xi_{0}=\dots=\xi_{n}=\xi}\\
\lambda_{1}\left(\xi\right) & =\left.\frac{\partial^{2}G}{\partial\xi_{\alpha}^{2}}\right|_{\xi_{0}=\dots=\xi_{n}=\xi}-\left.\frac{\partial^{2}G}{\partial\xi_{\alpha}\partial\xi_{\beta}}\right|_{\xi_{0}=\dots=\xi_{n}=\xi},
\end{align*}
where the first one is non-degenerate, whereas the second one is $n$-fold degenerate. Inserting eqs. (\ref{eq:HessGTc_diagonal}) and (\ref{eq:HessGTc_offdiagonal}), we obtain that, for $k=0$, 
\begin{align}
	\lambda_{0}\left(\xi\right) &\overset{k=0}{=} 0\\
	\lambda_{1}\left(\xi\right) &\overset{k=0}{=} -\left\langle \frac{\left[\Tc^{\prime}\left(\xi\right)\right]^{2}}{\Tc\left(\xi\right)\left(1-\Tc\left(\xi\right)\right)}\right\rangle_{r},\\ 
\end{align}
where the latter expression equals minus the Fisher information ${\cal I}_{\boldsymbol{n}}\left(\xi\right)$ for the stimulus $\xi$. Therefore, inserting this result into eq. (\ref{eq:RelMI_FI_with_EV}), we finally obtain
\begin{equation}
	I_{\mathrm{MI}} = \frac{1}{2}\left\langle \ln\left(\frac{N{\cal I}_{\boldsymbol{n}}\left(\xi\right)}{2\pi}\right)\right\rangle _{\xi\sim P_{\xi}}-\frac{1}{2}-\left\langle \ln\left(P_{\xi}\left(\xi\right)\right)\right\rangle _{\xi\sim P_{\xi}}+\mathcal{O}\left(\frac{1}{N}\right),
	\label{eq:RelMI_FI_with_FI}
\end{equation}
as expected according to \cite{Brunel98_1731}.



\begin{thebibliography}{0}%
\makeatletter
\providecommand \@ifxundefined [1]{%
 \@ifx{#1\undefined}
}%
\providecommand \@ifnum [1]{%
 \ifnum #1\expandafter \@firstoftwo
 \else \expandafter \@secondoftwo
 \fi
}%
\providecommand \@ifx [1]{%
 \ifx #1\expandafter \@firstoftwo
 \else \expandafter \@secondoftwo
 \fi
}%
\providecommand \natexlab [1]{#1}%
\providecommand \enquote  [1]{``#1''}%
\providecommand \bibnamefont  [1]{#1}%
\providecommand \bibfnamefont [1]{#1}%
\providecommand \citenamefont [1]{#1}%
\providecommand \href@noop [0]{\@secondoftwo}%
\providecommand \href [0]{\begingroup \@sanitize@url \@href}%
\providecommand \@href[1]{\@@startlink{#1}\@@href}%
\providecommand \@@href[1]{\endgroup#1\@@endlink}%
\providecommand \@sanitize@url [0]{\catcode `\\12\catcode `\$12\catcode
  `\&12\catcode `\#12\catcode `\^12\catcode `\_12\catcode `\%12\relax}%
\providecommand \@@startlink[1]{}%
\providecommand \@@endlink[0]{}%
\providecommand \url  [0]{\begingroup\@sanitize@url \@url }%
\providecommand \@url [1]{\endgroup\@href {#1}{\urlprefix }}%
\providecommand \urlprefix  [0]{URL }%
\providecommand \Eprint [0]{\href }%
\providecommand \doibase [0]{https://doi.org/}%
\providecommand \selectlanguage [0]{\@gobble}%
\providecommand \bibinfo  [0]{\@secondoftwo}%
\providecommand \bibfield  [0]{\@secondoftwo}%
\providecommand \translation [1]{[#1]}%
\providecommand \BibitemOpen [0]{}%
\providecommand \bibitemStop [0]{}%
\providecommand \bibitemNoStop [0]{.\EOS\space}%
\providecommand \EOS [0]{\spacefactor3000\relax}%
\providecommand \BibitemShut  [1]{\csname bibitem#1\endcsname}%
\let\auto@bib@innerbib\@empty
\end{thebibliography}%


\begin{thebibliography}{33}%
\makeatletter
\providecommand \@ifxundefined [1]{%
 \@ifx{#1\undefined}
}%
\providecommand \@ifnum [1]{%
 \ifnum #1\expandafter \@firstoftwo
 \else \expandafter \@secondoftwo
 \fi
}%
\providecommand \@ifx [1]{%
 \ifx #1\expandafter \@firstoftwo
 \else \expandafter \@secondoftwo
 \fi
}%
\providecommand \natexlab [1]{#1}%
\providecommand \enquote  [1]{``#1''}%
\providecommand \bibnamefont  [1]{#1}%
\providecommand \bibfnamefont [1]{#1}%
\providecommand \citenamefont [1]{#1}%
\providecommand \href@noop [0]{\@secondoftwo}%
\providecommand \href [0]{\begingroup \@sanitize@url \@href}%
\providecommand \@href[1]{\@@startlink{#1}\@@href}%
\providecommand \@@href[1]{\endgroup#1\@@endlink}%
\providecommand \@sanitize@url [0]{\catcode `\\12\catcode `\$12\catcode
  `\&12\catcode `\#12\catcode `\^12\catcode `\_12\catcode `\%12\relax}%
\providecommand \@@startlink[1]{}%
\providecommand \@@endlink[0]{}%
\providecommand \url  [0]{\begingroup\@sanitize@url \@url }%
\providecommand \@url [1]{\endgroup\@href {#1}{\urlprefix }}%
\providecommand \urlprefix  [0]{URL }%
\providecommand \Eprint [0]{\href }%
\providecommand \doibase [0]{https://doi.org/}%
\providecommand \selectlanguage [0]{\@gobble}%
\providecommand \bibinfo  [0]{\@secondoftwo}%
\providecommand \bibfield  [0]{\@secondoftwo}%
\providecommand \translation [1]{[#1]}%
\providecommand \BibitemOpen [0]{}%
\providecommand \bibitemStop [0]{}%
\providecommand \bibitemNoStop [0]{.\EOS\space}%
\providecommand \EOS [0]{\spacefactor3000\relax}%
\providecommand \BibitemShut  [1]{\csname bibitem#1\endcsname}%
\let\auto@bib@innerbib\@empty
\bibitem [{\citenamefont {Amari}(1977)}]{Amari77}%
  \BibitemOpen
  \bibfield  {author} {\bibinfo {author} {\bibfnamefont {S.}~\bibnamefont
  {Amari}},\ }\bibfield  {title} {\bibinfo {title} {Dynamics of pattern
  formation in lateral-inhibition type neural fields},\ }\href@noop {}
  {\bibfield  {journal} {\bibinfo  {journal} {Biological Cybernetics}\ }\textbf
  {\bibinfo {volume} {27}},\ \bibinfo {pages} {77} (\bibinfo {year}
  {1977})}\BibitemShut {NoStop}%
\bibitem [{\citenamefont {Ben-Yishai}\ \emph {et~al.}(1995)\citenamefont
  {Ben-Yishai}, \citenamefont {Bar-Or},\ and\ \citenamefont
  {Sompolinsky}}]{Benyishai95}%
  \BibitemOpen
  \bibfield  {author} {\bibinfo {author} {\bibfnamefont {R.}~\bibnamefont
  {Ben-Yishai}}, \bibinfo {author} {\bibfnamefont {R.~L.}\ \bibnamefont
  {Bar-Or}},\ and\ \bibinfo {author} {\bibfnamefont {H.}~\bibnamefont
  {Sompolinsky}},\ }\bibfield  {title} {\bibinfo {title} {Theory of orientation
  tuning in visual cortex},\ }\href@noop {} {\bibfield  {journal} {\bibinfo
  {journal} {Proceedings of the National Academy of Sciences of the United
  States of America}\ }\textbf {\bibinfo {volume} {92}} (\bibinfo {year}
  {1995})}\BibitemShut {NoStop}%
\bibitem [{\citenamefont {Tsodyks}\ and\ \citenamefont
  {Sejnowski}(1995)}]{Tsodyks95}%
  \BibitemOpen
  \bibfield  {author} {\bibinfo {author} {\bibfnamefont {M.}~\bibnamefont
  {Tsodyks}}\ and\ \bibinfo {author} {\bibfnamefont {T.}~\bibnamefont
  {Sejnowski}},\ }\bibfield  {title} {\bibinfo {title} {Associative memory and
  hippocampal place cells},\ }\href@noop {} {\bibfield  {journal} {\bibinfo
  {journal} {International journal of neural systems}\ }\textbf {\bibinfo
  {volume} {6}},\ \bibinfo {pages} {81} (\bibinfo {year} {1995})}\BibitemShut
  {NoStop}%
\bibitem [{\citenamefont {Battaglia}\ and\ \citenamefont
  {Treves}(1998)}]{Battaglia98}%
  \BibitemOpen
  \bibfield  {author} {\bibinfo {author} {\bibfnamefont {F.~P.}\ \bibnamefont
  {Battaglia}}\ and\ \bibinfo {author} {\bibfnamefont {A.}~\bibnamefont
  {Treves}},\ }\bibfield  {title} {\bibinfo {title} {Attractor neural networks
  storing multiple space representations: a model for hippocampal place
  fields},\ }\href@noop {} {\bibfield  {journal} {\bibinfo  {journal} {Physical
  Review E}\ }\textbf {\bibinfo {volume} {58}},\ \bibinfo {pages} {7738}
  (\bibinfo {year} {1998})}\BibitemShut {NoStop}%
\bibitem [{\citenamefont {Romani}\ and\ \citenamefont
  {Tsodyks}(2010)}]{romani10}%
  \BibitemOpen
  \bibfield  {author} {\bibinfo {author} {\bibfnamefont {S.}~\bibnamefont
  {Romani}}\ and\ \bibinfo {author} {\bibfnamefont {M.}~\bibnamefont
  {Tsodyks}},\ }\bibfield  {title} {\bibinfo {title} {Continuous attractors
  with morphed/correlated maps},\ }\href@noop {} {\bibfield  {journal}
  {\bibinfo  {journal} {PLoS computational biology}\ }\textbf {\bibinfo
  {volume} {6}},\ \bibinfo {pages} {e1000869} (\bibinfo {year} {2010})}\BibitemShut {NoStop}%
\bibitem [{\citenamefont {Battista}\ and\ \citenamefont
  {Monasson}(2020)}]{Battista20}%
  \BibitemOpen
  \bibfield  {author} {\bibinfo {author} {\bibfnamefont {A.}~\bibnamefont
  {Battista}}\ and\ \bibinfo {author} {\bibfnamefont {R.}~\bibnamefont
  {Monasson}},\ }\bibfield  {title} {\bibinfo {title} {Capacity-resolution
  trade-off in the optimal learning of multiple low-dimensional manifolds by
  attractor neural networks},\ }\href
  {https://doi.org/10.1103/PhysRevLett.124.048302} {\bibfield  {journal}
  {\bibinfo  {journal} {Phys. Rev. Lett.}\ }\textbf {\bibinfo {volume} {124}},\
  \bibinfo {pages} {048302} (\bibinfo {year} {2020})}\BibitemShut {NoStop}%
\bibitem [{\citenamefont {Posani}\ \emph {et~al.}(2018)\citenamefont {Posani},
  \citenamefont {Cocco},\ and\ \citenamefont {Monasson}}]{posani18}%
  \BibitemOpen
  \bibfield  {author} {\bibinfo {author} {\bibfnamefont {L.}~\bibnamefont
  {Posani}}, \bibinfo {author} {\bibfnamefont {S.}~\bibnamefont {Cocco}},\ and\
  \bibinfo {author} {\bibfnamefont {R.}~\bibnamefont {Monasson}},\ }\bibfield
  {title} {\bibinfo {title} {Integration and multiplexing of positional and
  contextual information by the hippocampal network},\ }\href@noop {}
  {\bibfield  {journal} {\bibinfo  {journal} {PLoS computational biology}\
  }\textbf {\bibinfo {volume} {14}},\ \bibinfo {pages} {e1006320} (\bibinfo
  {year} {2018})}\BibitemShut {NoStop}%
\bibitem [{\citenamefont {Yoon}\ \emph {et~al.}(2013)\citenamefont {Yoon},
  \citenamefont {Buice}, \citenamefont {Barry}, \citenamefont {Hayman},
  \citenamefont {Burgess},\ and\ \citenamefont {Fiete}}]{Yoon2013}%
  \BibitemOpen
  \bibfield  {author} {\bibinfo {author} {\bibfnamefont {K.}~\bibnamefont
  {Yoon}}, \bibinfo {author} {\bibfnamefont {M.~A.}\ \bibnamefont {Buice}},
  \bibinfo {author} {\bibfnamefont {C.}~\bibnamefont {Barry}}, \bibinfo
  {author} {\bibfnamefont {R.}~\bibnamefont {Hayman}}, \bibinfo {author}
  {\bibfnamefont {N.}~\bibnamefont {Burgess}},\ and\ \bibinfo {author}
  {\bibfnamefont {I.~R.}\ \bibnamefont {Fiete}},\ }\bibfield  {title} {\bibinfo
  {title} {{Specific evidence of low-dimensional continuous attractor dynamics
  in grid cells}},\ }\href
  {http://www.pubmedcentral.nih.gov/articlerender.fcgi?artid=3797513{\&}tool=pmcentrez{\&}rendertype=abstract}
  {\bibfield  {journal} {\bibinfo  {journal} {Nature Neuroscience}\ }\textbf
  {\bibinfo {volume} {16}},\ \bibinfo {pages} {1077} (\bibinfo {year}
  {2013})}\BibitemShut {NoStop}%
\bibitem [{\citenamefont {Wimmer}\ \emph {et~al.}(2014)\citenamefont {Wimmer},
  \citenamefont {Nykamp}, \citenamefont {Constantinidis},\ and\ \citenamefont
  {Compte}}]{Wimmer2014}%
  \BibitemOpen
  \bibfield  {author} {\bibinfo {author} {\bibfnamefont {K.}~\bibnamefont
  {Wimmer}}, \bibinfo {author} {\bibfnamefont {D.~Q.}\ \bibnamefont {Nykamp}},
  \bibinfo {author} {\bibfnamefont {C.}~\bibnamefont {Constantinidis}},\ and\
  \bibinfo {author} {\bibfnamefont {A.}~\bibnamefont {Compte}},\ }\bibfield
  {title} {\bibinfo {title} {Bump attractor dynamics in prefrontal cortex
  explains behavioral precision in spatial working memory},\ }\href@noop {}
  {\bibfield  {journal} {\bibinfo  {journal} {Nature neuroscience}\ }\textbf
  {\bibinfo {volume} {17}},\ \bibinfo {pages} {431} (\bibinfo {year}
  {2014})}\BibitemShut {NoStop}%
\bibitem [{\citenamefont {Kim}\ \emph {et~al.}(2017)\citenamefont {Kim},
  \citenamefont {Rouault}, \citenamefont {Shaul},\ and\ \citenamefont
  {Jayaraman}}]{Kim17_849}%
  \BibitemOpen
  \bibfield  {author} {\bibinfo {author} {\bibfnamefont {S.~S.}\ \bibnamefont
  {Kim}}, \bibinfo {author} {\bibfnamefont {H.}~\bibnamefont {Rouault}},
  \bibinfo {author} {\bibfnamefont {D.}~\bibnamefont {Shaul}},\ and\ \bibinfo
  {author} {\bibfnamefont {V.}~\bibnamefont {Jayaraman}},\ }\bibfield  {title}
  {\bibinfo {title} {Ring attractor dynamics in the drosophila central brain},\
  }\href {https://doi.org/10.1126/science.aal4835} {\bibfield  {journal}
  {\bibinfo  {journal} {Science}\ }\textbf {\bibinfo {volume} {356}},\ \bibinfo
  {pages} {849 } (\bibinfo {year} {2017})}\BibitemShut {NoStop}%
\bibitem [{\citenamefont {Seung}\ and\ \citenamefont
  {Sompolinsky}(1993)}]{Seung93_10749}%
  \BibitemOpen
  \bibfield  {author} {\bibinfo {author} {\bibfnamefont {H.~S.}\ \bibnamefont
  {Seung}}\ and\ \bibinfo {author} {\bibfnamefont {H.}~\bibnamefont
  {Sompolinsky}},\ }\bibfield  {title} {\bibinfo {title} {Simple models for
  reading neuronal population codes},\ }\href
  {https://doi.org/10.1073/pnas.90.22.10749} {\bibfield  {journal} {\bibinfo
  {journal} {Proceedings of the National Academy of Sciences}\ }\textbf
  {\bibinfo {volume} {90}},\ \bibinfo {pages} {10749} (\bibinfo {year}
  {1993})},\ \Eprint
  {https://arxiv.org/abs/https://www.pnas.org/content/90/22/10749.full.pdf}
  {https://www.pnas.org/content/90/22/10749.full.pdf} \BibitemShut {NoStop}%
\bibitem [{\citenamefont {Pouget}\ \emph {et~al.}(1999)\citenamefont {Pouget},
  \citenamefont {Deneve}, \citenamefont {Ducom},\ and\ \citenamefont
  {Latham}}]{Pouget99_85}%
  \BibitemOpen
  \bibfield  {author} {\bibinfo {author} {\bibfnamefont {A.}~\bibnamefont
  {Pouget}}, \bibinfo {author} {\bibfnamefont {S.}~\bibnamefont {Deneve}},
  \bibinfo {author} {\bibfnamefont {J.-C.}\ \bibnamefont {Ducom}},\ and\
  \bibinfo {author} {\bibfnamefont {P.~E.}\ \bibnamefont {Latham}},\ }\bibfield
   {title} {\bibinfo {title} {{Narrow Versus Wide Tuning Curves: What's Best
  for a Population Code?}},\ }\href
  {https://doi.org/10.1162/089976699300016818} {\bibfield  {journal} {\bibinfo
  {journal} {Neural Computation}\ }\textbf {\bibinfo {volume} {11}},\ \bibinfo
  {pages} {85} (\bibinfo {year} {1999})},\ \Eprint
  {https://arxiv.org/abs/https://direct.mit.edu/neco/article-pdf/11/1/85/814038/089976699300016818.pdf}
  {https://direct.mit.edu/neco/article-pdf/11/1/85/814038/089976699300016818.pdf}
  \BibitemShut {NoStop}%
\bibitem [{\citenamefont {Cover}\ and\ \citenamefont {Thomas}(1991)}]{Cover91}%
  \BibitemOpen
  \bibfield  {author} {\bibinfo {author} {\bibfnamefont {T.~M.}\ \bibnamefont
  {Cover}}\ and\ \bibinfo {author} {\bibfnamefont {J.~A.}\ \bibnamefont
  {Thomas}},\ }\href@noop {} {\emph {\bibinfo {title} {Elements of Information
  Theory}}}\ (\bibinfo  {publisher} {Wiley},\ \bibinfo {address} {New York},\
  \bibinfo {year} {1991})\BibitemShut {NoStop}%
\bibitem [{\citenamefont {Sompolinsky}\ \emph {et~al.}(2001)\citenamefont
  {Sompolinsky}, \citenamefont {Yoon}, \citenamefont {Kang},\ and\
  \citenamefont {Shamir}}]{Sompolinsky01_051904}%
  \BibitemOpen
  \bibfield  {author} {\bibinfo {author} {\bibfnamefont {H.}~\bibnamefont
  {Sompolinsky}}, \bibinfo {author} {\bibfnamefont {H.}~\bibnamefont {Yoon}},
  \bibinfo {author} {\bibfnamefont {K.}~\bibnamefont {Kang}},\ and\ \bibinfo
  {author} {\bibfnamefont {M.}~\bibnamefont {Shamir}},\ }\bibfield  {title}
  {\bibinfo {title} {Population coding in neuronal systems with correlated
  noise},\ }\href@noop {} {\bibfield  {journal} {\bibinfo  {journal} {Physical
  Review E}\ }\textbf {\bibinfo {volume} {64}},\ \bibinfo {pages} {051904}
  (\bibinfo {year} {2001})}\BibitemShut {NoStop}%
\bibitem [{\citenamefont {Brunel}\ and\ \citenamefont
  {Nadal}(1998)}]{Brunel98_1731}%
  \BibitemOpen
  \bibfield  {author} {\bibinfo {author} {\bibfnamefont {N.}~\bibnamefont
  {Brunel}}\ and\ \bibinfo {author} {\bibfnamefont {J.-P.}\ \bibnamefont
  {Nadal}},\ }\bibfield  {title} {\bibinfo {title} {Mutual information, fisher
  information, and population coding},\ }\href
  {https://doi.org/10.1162/089976698300017115} {\bibfield  {journal} {\bibinfo
  {journal} {Neural Computation}\ }\textbf {\bibinfo {volume} {10}},\ \bibinfo
  {pages} {1731} (\bibinfo {year} {1998})},\ \Eprint
  {https://arxiv.org/abs/https://doi.org/10.1162/089976698300017115}
  {https://doi.org/10.1162/089976698300017115} \BibitemShut {NoStop}%
\bibitem [{\citenamefont {Kriegeskorte}\ and\ \citenamefont
  {Wei}(2021)}]{Kriegeskorte21_703}%
  \BibitemOpen
  \bibfield  {author} {\bibinfo {author} {\bibfnamefont {N.}~\bibnamefont
  {Kriegeskorte}}\ and\ \bibinfo {author} {\bibfnamefont {X.-X.}\ \bibnamefont
  {Wei}},\ }\bibfield  {title} {\bibinfo {title} {Neural tuning and
  representational geometry},\ }\href
  {https://doi.org/10.1038/s41583-021-00502-3} {\bibfield  {journal} {\bibinfo
  {journal} {Nature Reviews Neuroscience}\ }\textbf {\bibinfo {volume} {22}},\
  \bibinfo {pages} {703 } (\bibinfo {year} {2021})}\BibitemShut {NoStop}%
\bibitem [{\citenamefont {Monasson}\ and\ \citenamefont
  {Rosay}(2013)}]{Rosay13_062813}%
  \BibitemOpen
  \bibfield  {author} {\bibinfo {author} {\bibfnamefont {R.}~\bibnamefont
  {Monasson}}\ and\ \bibinfo {author} {\bibfnamefont {S.}~\bibnamefont
  {Rosay}},\ }\bibfield  {title} {\bibinfo {title} {Crosstalk and transitions
  between multiple spatial maps in an attractor neural network model of the
  hippocampus: Phase diagram},\ }\href
  {https://doi.org/10.1103/PhysRevE.87.062813} {\bibfield  {journal} {\bibinfo
  {journal} {Phys. Rev. E}\ }\textbf {\bibinfo {volume} {87}},\ \bibinfo
  {pages} {062813} (\bibinfo {year} {2013})}\BibitemShut {NoStop}%
\bibitem [{\citenamefont {Wei}\ and\ \citenamefont
  {Stocker}(2016)}]{Wei16_305}%
  \BibitemOpen
  \bibfield  {author} {\bibinfo {author} {\bibfnamefont {X.-X.}\ \bibnamefont
  {Wei}}\ and\ \bibinfo {author} {\bibfnamefont {A.~A.}\ \bibnamefont
  {Stocker}},\ }\bibfield  {title} {\bibinfo {title} {{Mutual Information,
  Fisher Information, and Efficient Coding}},\ }\href
  {https://doi.org/10.1162/NECO_a_00804} {\bibfield  {journal} {\bibinfo
  {journal} {Neural Computation}\ }\textbf {\bibinfo {volume} {28}},\ \bibinfo
  {pages} {305} (\bibinfo {year} {2016})},\ \Eprint
  {https://arxiv.org/abs/https://direct.mit.edu/neco/article-pdf/28/2/305/955081/neco\_a\_00804.pdf}
  {https://direct.mit.edu/neco/article-pdf/28/2/305/955081/neco\_a\_00804.pdf}
  \BibitemShut {NoStop}%
\bibitem [{\citenamefont {M\'{e}zard}\ \emph {et~al.}(1987)\citenamefont
  {M\'{e}zard}, \citenamefont {Parisi},\ and\ \citenamefont
  {Virasoro}}]{Mezard87}%
  \BibitemOpen
  \bibfield  {author} {\bibinfo {author} {\bibfnamefont {M.}~\bibnamefont
  {M\'{e}zard}}, \bibinfo {author} {\bibfnamefont {G.}~\bibnamefont {Parisi}},\
  and\ \bibinfo {author} {\bibfnamefont {M.}~\bibnamefont {Virasoro}},\ }\href
  {http://www.worldcat.org/isbn/9971501163} {\emph {\bibinfo {title} {{Spin
  Glass Theory and Beyond (World Scientific Lecture Notes in Physics, Vol
  9)}}}}\ (\bibinfo  {publisher} {{World Scientific Publishing Company}},\
  \bibinfo {year} {1987})\BibitemShut {NoStop}%
\bibitem [{\citenamefont {Sherrington}\ and\ \citenamefont
  {Kirkpatrick}(1975)}]{Sherrington75_1792}%
  \BibitemOpen
  \bibfield  {author} {\bibinfo {author} {\bibfnamefont {D.}~\bibnamefont
  {Sherrington}}\ and\ \bibinfo {author} {\bibfnamefont {S.}~\bibnamefont
  {Kirkpatrick}},\ }\bibfield  {title} {\bibinfo {title} {Solvable model of a
  spin-glass},\ }\href@noop {} {\bibfield  {journal} {\bibinfo  {journal}
  {Physical Review Letters}\ }\textbf {\bibinfo {volume} {35}},\ \bibinfo
  {pages} {1792} (\bibinfo {year} {1975})}\BibitemShut {NoStop}%
\bibitem [{\citenamefont {K\"uhn}(2023)}]{Kuehn23_SpacentationCode}%
  \BibitemOpen
  \bibfield  {author} {\bibinfo {author} {\bibfnamefont {T.}~\bibnamefont
  {K\"uhn}},\ }\href {https://github.com/tobikausk/spacentation_code} {\emph
  {\bibinfo {title} {https://github.com/tobikausk/spacentation\_code}}}
  (\bibinfo {year} {2023})\BibitemShut {NoStop}%
\bibitem [{\citenamefont {Moser}\ \emph {et~al.}(2014)\citenamefont {Moser},
  \citenamefont {Roudi}, \citenamefont {Witter}, \citenamefont {Kentros},
  \citenamefont {Bonhoeffer},\ and\ \citenamefont {Moser}}]{Moser14_466}%
  \BibitemOpen
  \bibfield  {author} {\bibinfo {author} {\bibfnamefont {E.~I.}\ \bibnamefont
  {Moser}}, \bibinfo {author} {\bibfnamefont {Y.}~\bibnamefont {Roudi}},
  \bibinfo {author} {\bibfnamefont {M.~P.}\ \bibnamefont {Witter}}, \bibinfo
  {author} {\bibfnamefont {C.}~\bibnamefont {Kentros}}, \bibinfo {author}
  {\bibfnamefont {T.}~\bibnamefont {Bonhoeffer}},\ and\ \bibinfo {author}
  {\bibfnamefont {M.-B.}\ \bibnamefont {Moser}},\ }\bibfield  {title} {\bibinfo
  {title} {Grid cells and cortical representation},\ }\href
  {https://doi.org/10.1038/nrn3766} {\bibfield  {journal} {\bibinfo  {journal}
  {Nature Reviews Neuroscience}\ }\textbf {\bibinfo {volume} {15}},\ \bibinfo
  {pages} {466} (\bibinfo {year} {2014})}\BibitemShut {NoStop}%
\bibitem [{\citenamefont {Monasson}\ and\ \citenamefont
  {Rosay}(2014)}]{Monasson14_032803}%
  \BibitemOpen
  \bibfield  {author} {\bibinfo {author} {\bibfnamefont {R.}~\bibnamefont
  {Monasson}}\ and\ \bibinfo {author} {\bibfnamefont {S.}~\bibnamefont
  {Rosay}},\ }\bibfield  {title} {\bibinfo {title} {Crosstalk and transitions
  between multiple spatial maps in an attractor neural network model of the
  hippocampus: Collective motion of the activity},\ }\href
  {https://doi.org/10.1103/PhysRevE.89.032803} {\bibfield  {journal} {\bibinfo
  {journal} {Phys. Rev. E}\ }\textbf {\bibinfo {volume} {89}},\ \bibinfo
  {pages} {032803} (\bibinfo {year} {2014})}\BibitemShut {NoStop}%
\bibitem [{\citenamefont {Richter}\ and\ \citenamefont
  {Gjorgjieva}(2017)}]{Richter17_39}%
  \BibitemOpen
  \bibfield  {author} {\bibinfo {author} {\bibfnamefont {L.~M.}\ \bibnamefont
  {Richter}}\ and\ \bibinfo {author} {\bibfnamefont {J.}~\bibnamefont
  {Gjorgjieva}},\ }\bibfield  {title} {\bibinfo {title} {Understanding neural
  circuit development through theory and models},\ }\href
  {https://doi.org/https://doi.org/10.1016/j.conb.2017.07.004} {\bibfield
  {journal} {\bibinfo  {journal} {Current Opinion in Neurobiology}\ }\textbf
  {\bibinfo {volume} {46}},\ \bibinfo {pages} {39} (\bibinfo {year}
  {2017})}\BibitemShut {NoStop}%
\bibitem [{\citenamefont {Ko}\ \emph {et~al.}(2013)\citenamefont {Ko},
  \citenamefont {Cossell}, \citenamefont {Baragli}, \citenamefont {Antolik},
  \citenamefont {Clopath}, \citenamefont {Hofer},\ and\ \citenamefont
  {Mrsic-Flogel}}]{Ko13_96}%
  \BibitemOpen
  \bibfield  {author} {\bibinfo {author} {\bibfnamefont {H.}~\bibnamefont
  {Ko}}, \bibinfo {author} {\bibfnamefont {L.}~\bibnamefont {Cossell}},
  \bibinfo {author} {\bibfnamefont {C.}~\bibnamefont {Baragli}}, \bibinfo
  {author} {\bibfnamefont {J.}~\bibnamefont {Antolik}}, \bibinfo {author}
  {\bibfnamefont {C.}~\bibnamefont {Clopath}}, \bibinfo {author} {\bibfnamefont
  {S.~B.}\ \bibnamefont {Hofer}},\ and\ \bibinfo {author} {\bibfnamefont
  {T.~D.}\ \bibnamefont {Mrsic-Flogel}},\ }\bibfield  {title} {\bibinfo {title}
  {The emergence of functional microcircuits in visual cortex},\ }\href
  {https://doi.org/10.1038/nature12015} {\bibfield  {journal} {\bibinfo
  {journal} {Nature}\ }\textbf {\bibinfo {volume} {496}},\ \bibinfo {pages} {96
  } (\bibinfo {year} {2013})}\BibitemShut {NoStop}%
\bibitem [{\citenamefont {Sadeh}\ \emph {et~al.}(2015)\citenamefont {Sadeh},
  \citenamefont {Clopath},\ and\ \citenamefont {Rotter}}]{Sadeh15_1}%
  \BibitemOpen
  \bibfield  {author} {\bibinfo {author} {\bibfnamefont {S.}~\bibnamefont
  {Sadeh}}, \bibinfo {author} {\bibfnamefont {C.}~\bibnamefont {Clopath}},\
  and\ \bibinfo {author} {\bibfnamefont {S.}~\bibnamefont {Rotter}},\
  }\bibfield  {title} {\bibinfo {title} {Emergence of functional specificity in
  balanced networks with synaptic plasticity},\ }\href
  {https://doi.org/10.1371/journal.pcbi.1004307} {\bibfield  {journal}
  {\bibinfo  {journal} {PLOS Computational Biology}\ }\textbf {\bibinfo
  {volume} {11}},\ \bibinfo {pages} {1} (\bibinfo {year} {2015})}\BibitemShut
  {NoStop}%
\bibitem [{\citenamefont {Lim}(2019)}]{Lim19_e44098}%
  \BibitemOpen
  \bibfield  {author} {\bibinfo {author} {\bibfnamefont {S.}~\bibnamefont
  {Lim}},\ }\bibfield  {title} {\bibinfo {title} {Mechanisms underlying
  sharpening of visual response dynamics with familiarity},\ }\href
  {https://doi.org/10.7554/eLife.44098} {\bibfield  {journal} {\bibinfo
  {journal} {eLife}\ }\textbf {\bibinfo {volume} {8}},\ \bibinfo {pages}
  {e44098} (\bibinfo {year} {2019})}\BibitemShut {NoStop}%
\bibitem [{\citenamefont {Treves}(1990{\natexlab{a}})}]{Treves90_2631}%
  \BibitemOpen
  \bibfield  {author} {\bibinfo {author} {\bibfnamefont {A.}~\bibnamefont
  {Treves}},\ }\bibfield  {title} {\bibinfo {title} {Threshold-linear formal
  neurons in auto-associative nets},\ }\href
  {https://doi.org/10.1088/0305-4470/23/12/037} {\bibfield  {journal} {\bibinfo
   {journal} {Journal of Physics A: Mathematical and General}\ }\textbf
  {\bibinfo {volume} {23}},\ \bibinfo {pages} {2631} (\bibinfo {year}
  {1990}{\natexlab{a}})}\BibitemShut {NoStop}%
\bibitem [{\citenamefont {Treves}(1990{\natexlab{b}})}]{Treves90_2418}%
  \BibitemOpen
  \bibfield  {author} {\bibinfo {author} {\bibfnamefont {A.}~\bibnamefont
  {Treves}},\ }\bibfield  {title} {\bibinfo {title} {Graded-response neurons
  and information encodings in autoassociative memories},\ }\href
  {https://doi.org/10.1103/PhysRevA.42.2418} {\bibfield  {journal} {\bibinfo
  {journal} {Phys. Rev. A}\ }\textbf {\bibinfo {volume} {42}},\ \bibinfo
  {pages} {2418} (\bibinfo {year} {1990}{\natexlab{b}})}\BibitemShut {NoStop}%
\bibitem [{\citenamefont {Monasson}\ and\ \citenamefont
  {Rosay}(2015)}]{Monasson15_098101}%
  \BibitemOpen
  \bibfield  {author} {\bibinfo {author} {\bibfnamefont {R.}~\bibnamefont
  {Monasson}}\ and\ \bibinfo {author} {\bibfnamefont {S.}~\bibnamefont
  {Rosay}},\ }\bibfield  {title} {\bibinfo {title} {Transitions between spatial
  attractors in place-cell models},\ }\href
  {https://doi.org/10.1103/PhysRevLett.115.098101} {\bibfield  {journal}
  {\bibinfo  {journal} {Phys. Rev. Lett.}\ }\textbf {\bibinfo {volume} {115}},\
  \bibinfo {pages} {098101} (\bibinfo {year} {2015})}\BibitemShut {NoStop}%
\bibitem [{\citenamefont {Bourboulou}\ \emph {et~al.}(2019)\citenamefont
  {Bourboulou}, \citenamefont {Marti}, \citenamefont {Michon}, \citenamefont
  {El~Feghaly}, \citenamefont {Nouguier}, \citenamefont {Robbe}, \citenamefont
  {Koenig},\ and\ \citenamefont {Epsztein}}]{Bourboulou19_e44487}%
  \BibitemOpen
  \bibfield  {author} {\bibinfo {author} {\bibfnamefont {R.}~\bibnamefont
  {Bourboulou}}, \bibinfo {author} {\bibfnamefont {G.}~\bibnamefont {Marti}},
  \bibinfo {author} {\bibfnamefont {F.-X.}\ \bibnamefont {Michon}}, \bibinfo
  {author} {\bibfnamefont {E.}~\bibnamefont {El~Feghaly}}, \bibinfo {author}
  {\bibfnamefont {M.}~\bibnamefont {Nouguier}}, \bibinfo {author}
  {\bibfnamefont {D.}~\bibnamefont {Robbe}}, \bibinfo {author} {\bibfnamefont
  {J.}~\bibnamefont {Koenig}},\ and\ \bibinfo {author} {\bibfnamefont
  {J.}~\bibnamefont {Epsztein}},\ }\bibfield  {title} {\bibinfo {title}
  {Dynamic control of hippocampal spatial coding resolution by local visual
  cues},\ }\href {https://doi.org/10.7554/eLife.44487} {\bibfield  {journal}
  {\bibinfo  {journal} {eLife}\ }\textbf {\bibinfo {volume} {8}},\ \bibinfo
  {pages} {e44487} (\bibinfo {year} {2019})}\BibitemShut {NoStop}%
\bibitem [{\citenamefont {Touchette}(2009)}]{Touchette09_1}%
  \BibitemOpen
  \bibfield  {author} {\bibinfo {author} {\bibfnamefont {H.}~\bibnamefont
  {Touchette}},\ }\bibfield  {title} {\bibinfo {title} {The large deviation
  approach to statistical mechanics},\ }\href@noop {} {\bibfield  {journal}
  {\bibinfo  {journal} {Physics Reports}\ }\textbf {\bibinfo {volume} {478}},\
  \bibinfo {pages} {1} (\bibinfo {year} {2009})}\BibitemShut {NoStop}%
\bibitem [{\citenamefont {Nadal}\ and\ \citenamefont
  {Parga}(1993)}]{Nadal93_295}%
  \BibitemOpen
  \bibfield  {author} {\bibinfo {author} {\bibfnamefont {J.-P.}\ \bibnamefont
  {Nadal}}\ and\ \bibinfo {author} {\bibfnamefont {N.}~\bibnamefont {Parga}},\
  }\bibfield  {title} {\bibinfo {title} {Information processing by a perceptron
  in an unsupervised learning task},\ }\href
  {https://doi.org/10.1088/0954-898X\_4\_3\_004} {\bibfield  {journal}
  {\bibinfo  {journal} {Network: Computation in Neural Systems}\ }\textbf
  {\bibinfo {volume} {4}},\ \bibinfo {pages} {295} (\bibinfo {year} {1993})},\
  \Eprint {https://arxiv.org/abs/https://doi.org/10.1088/0954-898X\_4\_3\_004}
  {https://doi.org/10.1088/0954-898X\_4\_3\_004} \BibitemShut {NoStop}%
\end{thebibliography}

\providecommand{\noopsort}[1]{}\providecommand{\singleletter}[1]{#1}%

\providecommand{\noopsort}[1]{}\providecommand{\singleletter}[1]{#1}%

\end{document}